\documentclass[fleqn,usenatbib]{mnras}

\usepackage{newtxtext,newtxmath}

\usepackage[T1]{fontenc}

\DeclareRobustCommand{\VAN}[3]{#2}
\let\VANthebibliography\thebibliography
\def\thebibliography{\DeclareRobustCommand{\VAN}[3]{##3}\VANthebibliography}

\usepackage{fix-cm}
\usepackage{rotating}
\usepackage{caption}
\usepackage{subcaption}
\usepackage{booktabs}
\usepackage{multirow}
\usepackage[normalem]{ulem} 

\usepackage{graphicx}	
\usepackage{amsmath}	

\title[ML Identification of Infalling Groups]{Identifying group galaxies merging with massive clusters using machine learning}

\author[R. Jordan et al.]{Rhys Jordan,$^{1}$\thanks{E-mail: rhys.jordan@nottingham.ac.uk}
Meghan E. Gray,$^{1}$ 
Alfonso Arag\'{o}n-Salamanca,$^{1}$ 
Steven P. Bamford,$^{1}$ 
Frazer R. Pearce,$^{1}$
\newauthor
Roan Haggar$^{2}$
\\
$^{1}$School of Physics and Astronomy, University of Nottingham, Nottingham NG7 2RD, UK\\
$^{2}$Department of Physics and Astronomy, University of Waterloo, Waterloo, Ontario N2L 3G1, Canada\\
}

\date{Accepted XXX. Received YYY; in original form ZZZ}

\pubyear{\the\year{}}

\begin{document}
\label{firstpage}
\pagerange{\pageref{firstpage}--\pageref{lastpage}}
\maketitle

\begin{abstract}
The environment plays a critical role in galaxy evolution, with galaxy clusters and their infall regions offering diverse conditions that shape galaxies before they enter the dense cluster core, a process known as ``pre-processing''. However, identifying environmental substructures, particularly galaxy groups in these transitional zones, remains challenging due to projection effects and ``fingers-of-god'' distortions. In this work, we present a supervised machine learning framework for classifying galaxies into three environmental categories: \textit{main cluster}, \textit{group}, and \textit{neither}, using observable galaxy properties such as positions, line-of-sight velocities, and stellar mass. The model is trained on mock observations derived from cosmological simulations designed to replicate survey conditions and achieves an overall accuracy and class-size-weighted precision of 81\%. The \textit{neither}-type and \textit{Main cluster} galaxies are reliably recovered, whereas \textit{group} galaxies remain the most challenging to identify, achieving 30\% completeness and 76\% purity. Within $1\times R_{200}$ classification performance is suppressed, but it improves beyond this radius, reaching 40\% completeness and 80\% purity. Resampling and thresholding strategies allow the model to be tuned toward either higher purity or higher completeness; in this study, we adopt first-past-the-post thresholding to emphasise purity. Model performance is consistent across cluster masses and dynamical states, and it outperforms both Friends-of-Friends and Gaussian Mixture Modelling. This flexibility makes it well-suited to upcoming spectroscopic surveys of cluster infall regions, providing a robust statistical tool for disentangling environmental influences on galaxy evolution.
\end{abstract}



\section{Introduction}\label{sec: introduction}

Galaxy clusters are the most massive gravitationally bound structures in the universe, forming hierarchically from primordial density fluctuations and marking the nodes of the cosmic web \citep{White:1978, Davis:1985, Springel:2006}. They grow through the accretion of galaxies along with their dark matter haloes, interstellar medium, stars and dust. Accretion pathways range from infall of galaxies from the cosmic web to complex mergers. Galaxy accretion onto clusters via cosmic filaments is the most common process \citep{Colberg:1999, Mann:2012, Castignani:2022}, along with galaxy groups \citep{McGee:2009}, with filaments hosting up to 45\% of galaxies in the infall region around rich clusters at $z\simeq0$ \citep{Kuchner:2022}. Galaxies consequently experience a diverse range of environments in which galaxy morphology and star formation activity are known to be strongly environment-dependent. This is encapsulated in the well-established morphology–density \citep{Dressler:1980, Postman:1984} and star formation rate (SFR)–density \citep{Balogh:1998} relations, which show that dense environments, such as groups and clusters, predominantly host early-type, red, and quiescent galaxies \citep{Oemler:1974, Davis:1976}. In contrast, late-type, blue, star-forming galaxies are more frequently found in lower-density regions such as the field and voids \citep{Lewis:2002, Gomez:2003, Kauffmann:2004}. This suggests that higher-density environments lead to the suppression of star formation, known as quenching. The impact of the environment was found to be greater on SFR than on morphology \citep{Kauffmann:2004, Ball:2008, Bamford:2009}.

Despite significant progress, the precise mechanisms responsible for galaxy quenching in different environments remain poorly understood. Proposed external mechanisms include interactions with the intra-group and intra-cluster medium, such as ram pressure stripping \citep{Gunn:1972, Abadi:1999}, the suppression of gas accretion due to the host cluster’s deep gravitational potential (‘starvation’ or ‘strangulation’; \citep{Larson:1980}), repeated high-speed encounters with neighbouring galaxies (‘harassment’; \citealt{Moore:1996, Park:2009}), and tidal effects from the cluster potential, which can strip or stir galaxy mass \citep{Valluri:1993, Gnedin:2003}. Internal mechanisms, such as feedback from AGN \citep{Best:2005, Hopkins:2005} or supernovae \citep{Dekel:1986, Dekel:2003}, may also play a critical role.

There remains debate over whether evolution results from nature or nurture — whether a galaxy’s evolution is driven by intrinsic properties established early in its formation (\textit{nature}) or external environmental factors (\textit{nurture}). Galaxy groups are a key element in addressing this question. Groups rarely survive infall into clusters, as their member galaxies are tidally disrupted or unbound under the influence of the cluster’s deep gravitational potential \citep{Haggar:2023}. The group and its environment are disrupted during the first pericentric passage, as galaxies transition into cluster substructures, dynamically and spatially coherent collections of galaxies within the cluster that are not yet virialised with respect to the main halo. Consequently, groups serve as valuable laboratories for studying galaxy properties and morphology changes before direct exposure to the extreme conditions of the cluster core, referred to as “pre-processing” \cite{Wetzel:2012}. As groups approach the non-linear regime of gravitational collapse, typically defined as the infall region ($2$-$4 \times R_{200}$; \citealt{Diaferio:1997}), the distinction between group and cluster environments becomes increasingly ambiguous, introducing complications to those studying pre-processing. 

To investigate how the environment influences galaxy evolution, one must first accurately classify the environment in which a galaxy resides. Many studies have tackled this challenge, including efforts to identify filament galaxies using both classical techniques \citep{Cornwell:2023} and machine learning approaches \citep{Weaver:2023}, as well as to assign cluster membership through deep learning models incorporating photometric and spectroscopic data \citep{Angora:2020, Hashimoto:2024}. However, the identification of galaxy groups, and by extension, cluster substructures, has traditionally relied on spatial clustering algorithms, such as the friends-of-friends (FoF) method \citep{Huchra:1982, Davis:1985, Duarte:2014, Tempel:2016, Rodriguez:2020}. In the vicinity of massive clusters and within the clusters themselves, alternative methods, such as the Dressler–Shectman test \citep{Dressler:1988} and its successors \citep{Bird:1994, Girardi:1997, Ferrari:2003, Girardi:2015, Benavides:2023}, incorporate velocity information to identify local kinematic deviations. However, a key limitation of the Dressler–Shectman test is its inability to assign galaxies to specific substructures.

More recently, efforts have been made to combine spatial and velocity information, typically accessible through spectroscopic surveys, to define galaxy groups in three-dimensional phase space (2D position + 1D line-of-sight velocity) \citep{Serna:1996, Yu:2015, Yu:2018, Lavaux:2016, Lorenco:2020, Monteiro-Oliveira:2022, Piraino-Cerda:2024}. Despite their utility, clustering algorithms such as Gaussian Mixture Models often suffer from interpretability issues and rely on heuristic criteria for model selection, making their results less definitive than those of supervised approaches.

In this work, we pursue a more direct and robust alternative: leveraging cosmological simulations of large-scale structure, constructed to match observational constraints, to train supervised machine learning (ML) models for allocating galaxies to their environments. The capacity of supervised ML to learn complex, non-linear mappings between observational features makes it particularly well-suited for this task, especially given the distortions introduced by projection effects and redshift-space phenomena such as the fingers-of-god (FoG; \citealt{Tully:1978}) and the Kaiser effect \citep{Kaiser:1987}, which pose a significant challenge in the cluster non-linear infall region (within $\sim15~h^{-1}$~Mpc; \citealt{Kuchner:2021}). Our classification is based on the intrinsic three-dimensional distribution of galaxies and their full halo-to-halo gravitational boundness (see Section~\ref{subsec: substructure catalogues}). Trained on a large and representative sample of simulated clusters, our method is readily applicable to existing wide-area spectroscopic surveys, including SDSS redMaPPer \citep{Rykoff:2014}, GAMA \citep{Robotham:2011}, and DESI \citep{Marini:2025}, as well as to forthcoming wide-field cluster surveys such as WEAVE \citep{Jin:2023} and CHANCES \citep{Sifon:2025}. Its application will support detailed dynamical and structural analyses of the cluster environment and residing galaxies. While our primary model includes the stellar mass of the target galaxy as an input feature, we also explore a variant in which this information is excluded, allowing us to assess the extent to which environmental classification can be achieved independently of intrinsic galaxy properties.

This paper is the first in a two-part series. In Paper I, we present our method for assigning galaxies to broad environmental classes. The companion paper II will associate individual galaxies with discrete groups and characterise the group properties. The structure of this paper is as follows: Section~\ref{sec: data and catalogues} introduces the \textsc{TheThreeHundred} simulation and mock observation datasets. Section~\ref{sec: supervised machine learning} outlines our supervised machine learning classification methodology. This is followed by Section~\ref{sec: results}, which presents and simultaneously discusses our results. Finally, Section~\ref{sec: conclusions} summarises our main findings.   

\section{Data and Catalogues}\label{sec: data and catalogues}

\subsection{Cosmological Simulations}

Our data is sourced from the \textsc{TheThreeHundred} project \citep{Cui:2018}, which provides hydrodynamical re-simulations of the 324 most massive galaxy clusters identified in the dark matter-only MultiDark Planck 2 (MDPL2) simulation \citep{Klypin:2016}. MDPL2 evolves the formation of large-scale structure in a $1~h^{-1}\mathrm{Gpc}$ comoving cube, using cosmological parameters from \citealt{Planck_Collaboration:2016}. The simulation contains $3840^{3}$ dark matter particles, each with a mass of $1.5 \times 10^{9}~\mathrm{M}_{\odot}$. Haloes were identified using the \textsc{rockstar} halo finder, and the 324 most massive were selected as targets for zoom resimulations \citep{Behroozi:2013}.

\textsc{TheThreeHundred} then zooms in on these clusters by centring a $15~h^{-1}\mathrm{Mpc}$ comoving cube on each most massive halo. Original MDPL2 dark matter particles within this volume are traced back to their origin and split into dark matter and gas particles with masses of $m_{\mathrm{DM}}=1.27 \times 10^{9}~h^{-1}\mathrm{M}_{\odot}$ and $m_{\mathrm{gas}}=2.36 \times 10^{8}~h^{-1}\mathrm{M}_{\odot}$, respectively. Lower-resolution particles are placed outside the high-resolution region to preserve tidal forces from the surrounding large-scale structure.

These zoom-in regions are then re-simulated using hydrodynamics across 128 snapshots spanning the redshift range $z=17$ to $z=0$. In this study, we use the \textsc{gadget-x} run, a modernised implementation of the smoothed particle hydrodynamics code \textsc{Gadget-2} \citep{Springel:2005}, which includes improved baryonic physics and feedback prescriptions.

\subsection{Substructure catalogues}\label{subsec: substructure catalogues}

The halo catalogues used in this study were generated using Amiga’s Halo Finder (AHF; \citealt{Knollmann:2009}). AHF identifies haloes as spherical over-densities and iteratively unbinds particles from the outskirts inward until a self-bound structure remains, requiring a minimum of 20 bound particles to retain a halo. All associated mass is included in the total halo mass for parent haloes hosting subhaloes.

Since our objective is to identify group galaxies, we must clearly distinguish them from galaxies in other environments. We categorise galaxies in three ways: those bound to the main cluster halo, those bound to group host haloes, and those not associated with any significant structure, referred to as `\textit{neither}'. The group galaxy catalogue was constructed following the definition in \cite{Han:2018}. A halo qualifies as a group host if it has at least one less massive halo bound to it. The boundness criterion for group members is defined by Equation (1) in \cite{Haggar:2023}, which requires the sum of a subhalo’s kinetic and potential energy to be less than the gravitational potential of the group host at a distance of $2.5 \times R_{200}^{\mathrm{grp}}$, assuming a Navarro-Frenk-White dark matter halo density profile. This distance reflects the approximate upper limit for backsplash galaxies \citep{Mamon:2004}. This is important as we are interested in galaxies that have experienced the group environment, not just those that are actively experiencing it. 

The halo binding scheme is exclusive and hierarchical: each halo may be bound only to a single, more massive halo, ensuring a unique association. The main cluster galaxy catalogue was constructed using this algorithm but applied exclusively to the main cluster halo, effectively treating it as a group host. Group catalogues were generated in parallel with the main cluster halo removed. This step was necessary because, under the exclusivity rule, group hosts bound to the main cluster halo, and consequently their member galaxies, would otherwise be classified as cluster galaxies. Removing the main cluster halo overcomes this limitation and permits the identification of groups within clusters. Each galaxy in the cluster vicinity is assigned membership according to whether it is more strongly bound to the main cluster or to a group host.

While the group catalogues retain galaxy membership within individual groups for each cluster, in this analysis, all group galaxies are treated collectively as a single group class, without distinguishing between specific groups. Groups are further restricted to a richness range of 5–50 members, including the host galaxy \citep{Tully:2015}. The lower limit excludes pairs and low-mass groups, whose environments are unlikely to exert a significant influence on their member galaxies \citep{Hester:2006}, while the upper limit avoids systems undergoing major cluster–cluster mergers, ensuring that only genuine group-scale structures are included.

Throughout this study, each halo is assumed to trace a galaxy; therefore, we use the term `galaxy' throughout.

\subsection{Dataset pre-processing}

Several issues needed to be resolved before the data could be used effectively. First, 20 clusters (10 pairs) overlapped within their respective re-simulation volumes, introducing potential duplication and more complicated infall scenarios within the AHF halo catalogues and the identification of substructure. These overlapping clusters were deemed unreliable and subsequently removed from the analysis.

Second, the mass resolution of the simulation imposed practical constraints. The total halo mass function was observed to turn over at approximately $10^{11}~m_\odot$, which is unphysical; structure formation in a hierarchical universe should produce an increasing number of lower-mass haloes. This turnover sets our lower halo mass limit. On the high end, galaxy-scale haloes rarely exceed $10^{13}~\mathrm{M}_{\odot}$, which we use as an upper limit. We further imposed a minimum group host halo mass of $10^{13.5}~m_\odot$. This threshold is slightly above the turnover in the group host mass function.

In observational studies, stellar mass is often preferred over halo mass as a proxy for galaxy properties, as it can be robustly estimated from photometric or spectroscopic data (e.g., \citealt{Bell:2001}). In contrast, the total halo mass is not easily measurable from observations. However, the \textsc{Gadget-X} simulations used in this work are known to produce unreliable stellar mass estimates for galaxies. To address this, we adopt the \textsc{UniverseMachine} framework \citep{Behroozi:2019} to calculate stellar masses for a given peak total halo mass. \citealt{Behroozi:2019} maps peak halo mass to stellar mass in a manner that statistically reproduces the observed stellar mass function across redshift, thereby offering a more realistic and observationally consistent connection between halos and galaxies.

Lastly, low-resolution particles were found to encroach significantly into some cluster regions, violating their intended Lagrangian boundaries and compromising the reliability of hydrodynamical properties. To preserve the integrity of the training data, we restricted the galaxy sample used for training, validation, and testing to within $7.5~\mathrm{Mpc}$ of the cluster centre, as low-resolution contamination is negligible within this radius for this sample. However, to avoid introducing artificial boundary effects when computing environment-dependent features, such as nearest-neighbour metrics, we retained galaxy data out to $15~\mathrm{Mpc}$ for feature generation (see Section~\ref{subsec: feature space}).

\begin{figure}
    \centering
    \includegraphics[width=1\linewidth]{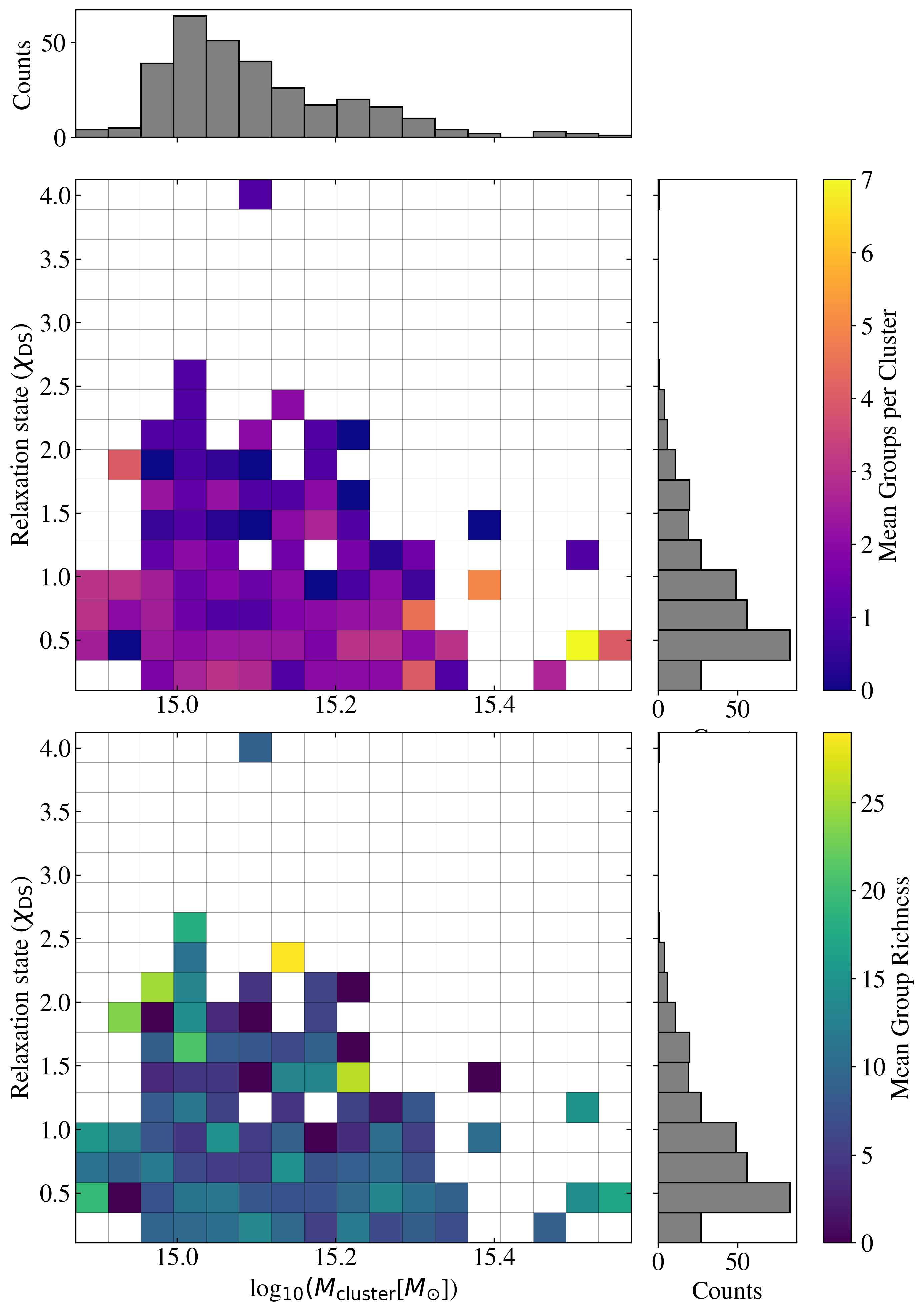}
    \caption{Group statistics across the cluster mass–relaxation plane. Each panel displays a binned 2D distribution overlaid with marginal histograms on top and to the right that represent the total number of clusters in each bin. \textit{Top}: Distribution of mean group counts per cluster as a function of total cluster mass and dynamical state. The majority of detected groups and their member galaxies reside in low-mass, dynamically disturbed clusters. This trend reflects population statistics — such clusters dominate the sample, rather than an intrinsic physical correlation. However, more massive, disturbed clusters are group-rich. \textit{Bottom}: Mean group richness as a function of cluster stellar mass and dynamical state. While group numbers are highest in low-mass, disturbed clusters, richer groups (on average) seem to preferentially reside in more dynamically relaxed systems.}
    \label{fig:cluster-group-stats}
\end{figure}

Our final sample consists of 304 simulated regions centred on clusters with a broad range of masses $14.9 \leq \log_{10}\!\left( M_{\mathrm{cluster}} \,[M_{\odot}] \right) \leq 15.6$) and dynamical state $(0.1 < \chi_{\mathrm{DS}} < 4.1)$, where lower $\chi_{\mathrm{DS}}$ values indicate more dynamically disturbed systems (see Haggar et al. 2020 and references therein for details on this parameter). These regions contain 300,781 galaxies spanning a stellar mass range of $8.56 \leq \log_{10} (M_{\ast}\,[M_\odot]) \leq 12.0$. Of these, 56,513 are cluster members, 18,148 group galaxies, and 226,120 `neither' galaxies. Of these, 258 clusters contain at least one group after applying our pre-processing criteria, yielding a total of 603 groups. Most groups are found in lower-mass, dynamically disturbed clusters, a reflection of population statistics, as these clusters numerically dominate the sample. Figure~\ref{fig:cluster-group-stats} illustrates how group statistics vary across cluster mass and relaxation state. Group-rich clusters tend to be both massive and disturbed; however, average group richness correlates more strongly with relaxation than mass, with richer groups being preferentially found in more relaxed clusters. We retain clusters without associated groups in our sample to serve as negative training examples and as a reference baseline of group-free systems.

\section{Supervised Machine Learning}\label{sec: supervised machine learning}

Machine learning (ML) is a data-driven approach that identifies patterns in input data during training and generates predictions when presented with new, unseen data. Depending on the nature of the output, ML tasks are typically framed as either classification (discrete outputs) or regression (continuous outputs) problems. A fundamental distinction in ML lies between supervised and unsupervised learning. Supervised learning involves training on labelled data, where the true outputs are known. This enables the model to learn a direct mapping from inputs to outputs, allowing for straightforward evaluation on unseen data. In contrast, unsupervised learning operates without ground truth labels, relying instead on alternative strategies to assess performance, such as clustering validity metrics or reconstruction errors. In this work, we focus exclusively on supervised ML.

Supervised ML workflows typically divide the dataset into training, validation, and test sets. The model is trained on the training set, while the validation set is used to tune hyperparameters and monitor performance, thereby mitigating over-fitting. Final performance assessment is conducted on the test set, which remains unseen during training and validation, ensuring an unbiased evaluation of the model's generalisation capability.

\subsection{Random Forests}

A Random Forest (RF) \citep{Breiman:2001} is a supervised ensemble learning method that constructs multiple decision trees in parallel. Each tree is composed of a sequence of binary decisions based on input features, recursively partitioning the data using thresholds that best separate the target classes. These splits are determined by minimising a measure of data impurity at each node. Two common impurity metrics are Gini impurity, which is computationally efficient, and Shannon entropy, which involves logarithmic calculations. In this study, we adopt Gini impurity as it provides comparable classification performance to entropy while offering significantly faster computation, particularly advantageous when training large ensembles. Each decision tree is grown by selecting the feature and threshold that yield the greatest reduction in impurity, and classification is determined when a terminal node reaches sufficient class purity.

In an RF, each tree is trained on a bootstrap sample (i.e., a randomly sampled subset with replacement) of the given dataset. This process enables the model to explore different regions of parameter space, mitigates overfitting, and reduces data variance. Votes are aggregated for all models, where the proportion of votes assigned to each class is interpreted as a class assignment probability. The final assignment is made to the class with the highest probability.

Compared to other ML algorithms, RFs are robust to complex, non-linear relationships, require minimal pre-processing, and are relatively insensitive to feature scaling. They include built-in validation tools, such as out-of-bag error estimation, and naturally provide feature importance rankings, which are key advantages when interpretability is essential. Furthermore, they respond well to rebalanced datasets through sampling techniques or class weight adjustments, making them well-suited for imbalanced classification problems. For these reasons, RFs were chosen as the ML model for this study.

We used a 60–20–20 training–validation–test split for our cluster datasets. To ensure the subsets, created by randomly selecting clusters, were statistically representative of the full sample, we applied two-sample Kolmogorov–Smirnov tests to the cluster mass and relaxedness distributions. In both cases, the resulting $p$-values were always above 0.15, indicating no significant differences and confirming that the subsets are statistically compatible with the overall distribution.

\subsection{Feature space}\label{subsec: feature space}

Classifying a target galaxy as part of a group relies on the properties of nearby galaxies, collectively deemed the environment. We use features in galaxy phase space, including each galaxy’s position, velocity, and stellar mass, all evaluated at redshift $z = 0$. To better characterise the environment, we extend this feature space by incorporating differences in these properties relative to the target’s $n$ nearest neighbours. This allows us to capture environmental information across a broad range of scales. Specifically, we adopt $n = 2, 3, 5, 10, 20, 30,$ and $50$, where $n=1$ refers to the target galaxy itself. In total, our feature space comprises 24 parameters per galaxy, computed first using full 3D simulation data and then using 2D projected mock observations.

\subsubsection{Simulation – 3D}

The first two features are the target galaxy’s cluster-centric [1] radial distance, $r_{3\mathrm{D}}$, and [2] radial velocity, $v_{r,3\mathrm{D}}$. Here, the radial velocity is the component of the 3D velocity vector in the direction of the cluster centre, $\hat{\mathbf{r}}$, and is computed via the dot product:
\begin{equation}\label{eqn:radial_velocity} 
    v_{r,3\mathrm{D}} = \mathbf{v} \cdot \hat{\mathbf{r}} = \frac{\mathbf{v} \cdot \mathbf{r}}{|\mathbf{r}|}, 
\end{equation}
where $\mathbf{v}$ and $\mathbf{r}$ are the galaxy’s 3D velocity and position vectors, respectively.

The next two sets of parameters describe the target galaxy’s local environment relative to its $n^{\mathrm{th}}$ nearest-neighbour: [3–9] the separation in radial distance, $\Delta r_{3\mathrm{D},n}$, and [10–16] the relative velocity, defined as the magnitude of the difference between their 3D velocity vectors, $\Delta v_{3\mathrm{D},n}$,
\begin{equation}\label{eqn:velocity_diff}
    \Delta v_{3\mathrm{D},n} = || \mathbf{v}_n - \mathbf{v}_1||
\end{equation}
where $\mathbf{v}_1$ is the velocity of the target galaxy and $\mathbf{v}_n$ is that of its $n^{\mathrm{th}}$ nearest-neighbour.

We also calculate [17–23] the local mass-weighted density, $\rho_{n}$, and is defined as the sum of stellar masses of the $n$ nearest neighbours divided by the volume of a sphere whose radius is the separation to the $n$th neighbour: \begin{equation}\label{eqn: mass-weighted local density 3D} 
    \rho_{n} = \frac{\sum_{i=1}^{n} m_{i}}{\frac{4}{3} \pi r_{3\mathrm{D},n}^3},
\end{equation} 
where $m_i$ is the stellar mass of the $i^{\mathrm{th}}$ neighbour. The final feature [24] is the stellar mass of the target galaxy, $m_{\ast}$.

\subsubsection{Observation – 2D}\label{sec: 2D obs FS}

To create mock observations, each cluster is projected along the line-of-sight (LoS) for all three coordinate axes, producing independent datasets for the $x$–$y$, $x$–$z$, and $y$–$z$ planes. This projection effectively reduces dimensionality by removing one spatial and two velocity components, resulting in a 2D analogue of the original feature space. 

We retain the same 24 features, but reinterpret them in the 2D observational context. The [1] radial distance, $r_{2\mathrm{D}}$, is now the projected 2D distance from the galaxy to the cluster centre. The [2] radial velocity, $v_{r,2\mathrm{D}}$, is the velocity component along the LoS, orthogonal to the projection plane. Nearest-neighbour separations in [3–9] radial distance, $\Delta r_{2\mathrm{D},n}$, correspond to the 2D sky-plane separation, while the [10–16] velocity differences, $\Delta v_{r,2\mathrm{D},n}$, are calculated from the difference in LoS velocities between the target and each neighbour. We prefer using velocity differences over velocity dispersion due to the noise sensitivity of dispersion estimates in low-count regimes, as noted in \citealt{Beers:1990}.

The [17–23] local density must be derived as a mass-weighted surface density normalised over the area of a circle defined by the sky distance to the $n$th nearest-neighbour. We adopt a more observationally motivated notation for this quantity used in \citealt{Wolf:2005, Wolf:2009},
\begin{equation}\label{eqn: mass-weighted local density 2D}
    \Sigma_{n}^{\ast} = \frac{\sum_{i=1}^{n} m_{i}}{\pi r_{2\mathrm{D},n}^2}. 
\end{equation}
Finally, the [24] stellar mass of the target galaxy remains unchanged as an intrinsic property.

The resulting distributions of feature-space values for simulation and observation are shown in Appendix~\ref {sec: appendix A}.

\section{Results}\label{sec: results}

\subsection{3D Simulation}

\begin{figure}
    \centering
    \includegraphics[width=\linewidth]{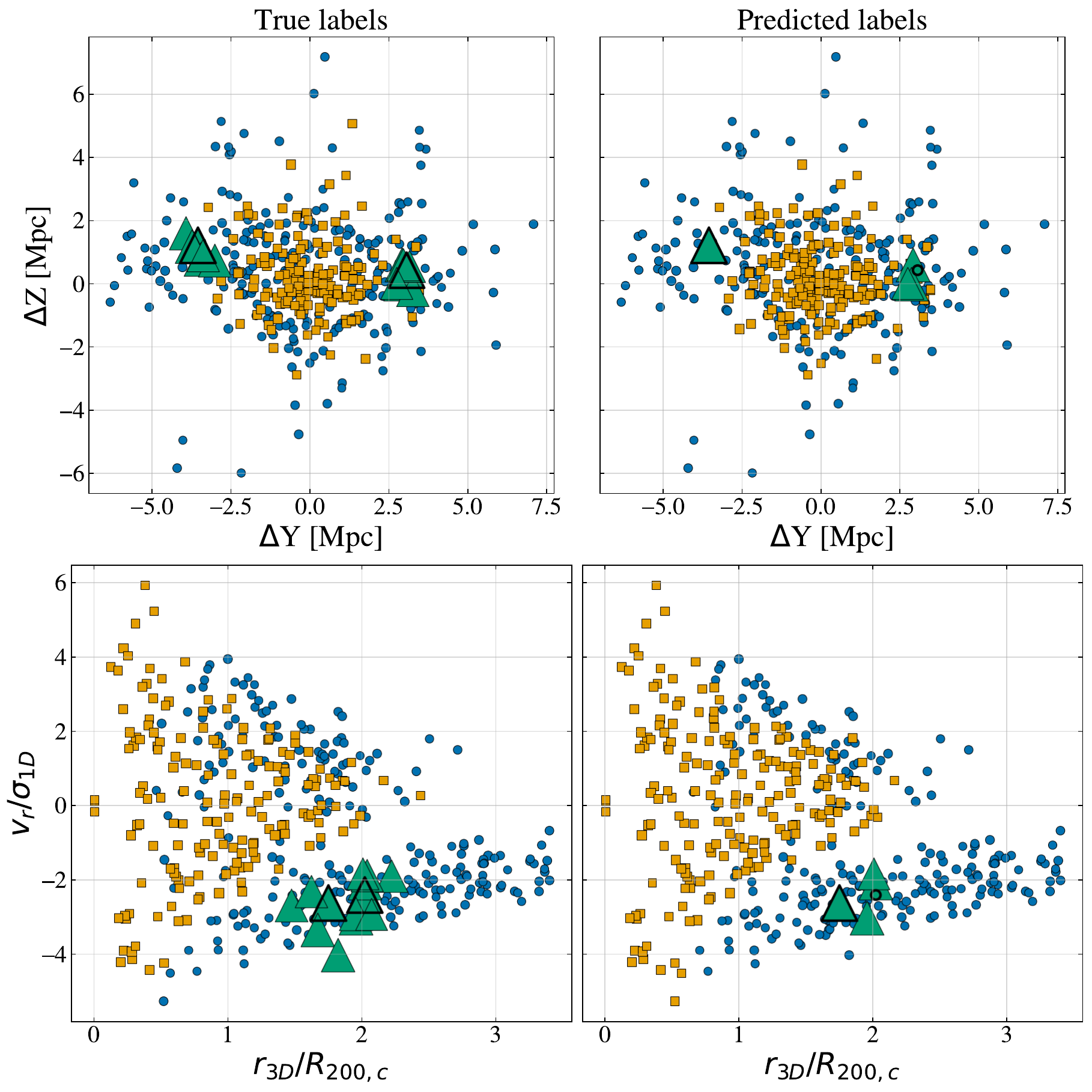}
        \caption{Cluster 155 plotted on the sky (\textit{top}) and in 3D radial velocity phase-space (\textit{bottom}). The left column shows the true galaxy classifications, indicated by marker shape and colour: blue circles for \textit{neither}, orange squares for \textit{main cluster}, and green triangles for \textit{group}. The right column shows the corresponding model predictions, using the same visual conventions. Phase-space velocities correspond to the 3D radial component of the galaxy's velocity towards the cluster centre, as defined in equation~\ref{eqn:radial_velocity}. Cluster velocity dispersion is derived from the halo $R_{200}$ in the AHF catalogue and scaled using equation~(8) from \protect\citealt{Finn:2005}, assuming standard Planck cosmology. Since only one velocity component is used, the 3D dispersion is converted to 1D by division by $\sqrt{3}$ (for further explanation on this type of 6D phase-space, refer to Figure~2 in \protect\citealt{Arthur:2019}). Group hosts (i.e., central group galaxies) are further highlighted with bold black outlines across all panels. The group host on the right-hand side of the cluster has been incorrectly assigned to the \textit{neither} class.}
    \label{fig: cluster 155 yz plane 3d}
\end{figure}

Figure~\ref{fig: cluster 155 yz plane 3d} presents an example of the sky-plane and phase-space predictions cluster 155, obtained using a Random Forest model trained on the full 3D information. Cluster 155 has an intermediate halo mass of $10^{15.1}$~M$_{\odot}$ compared to the rest of the simulation clusters and likewise with its dynamical state value of $\chi_{DS}=1.15$. While the model does not recover many group galaxies, which are a small minority of the total galaxy population, it identifies group hosts. The correctly identified group members are typically located near their corresponding hosts. In contrast, the \textit{main cluster} and \textit{neither} populations are recovered with relatively high fidelity.

\begin{figure}
    \centering
    \includegraphics[width=1\linewidth]{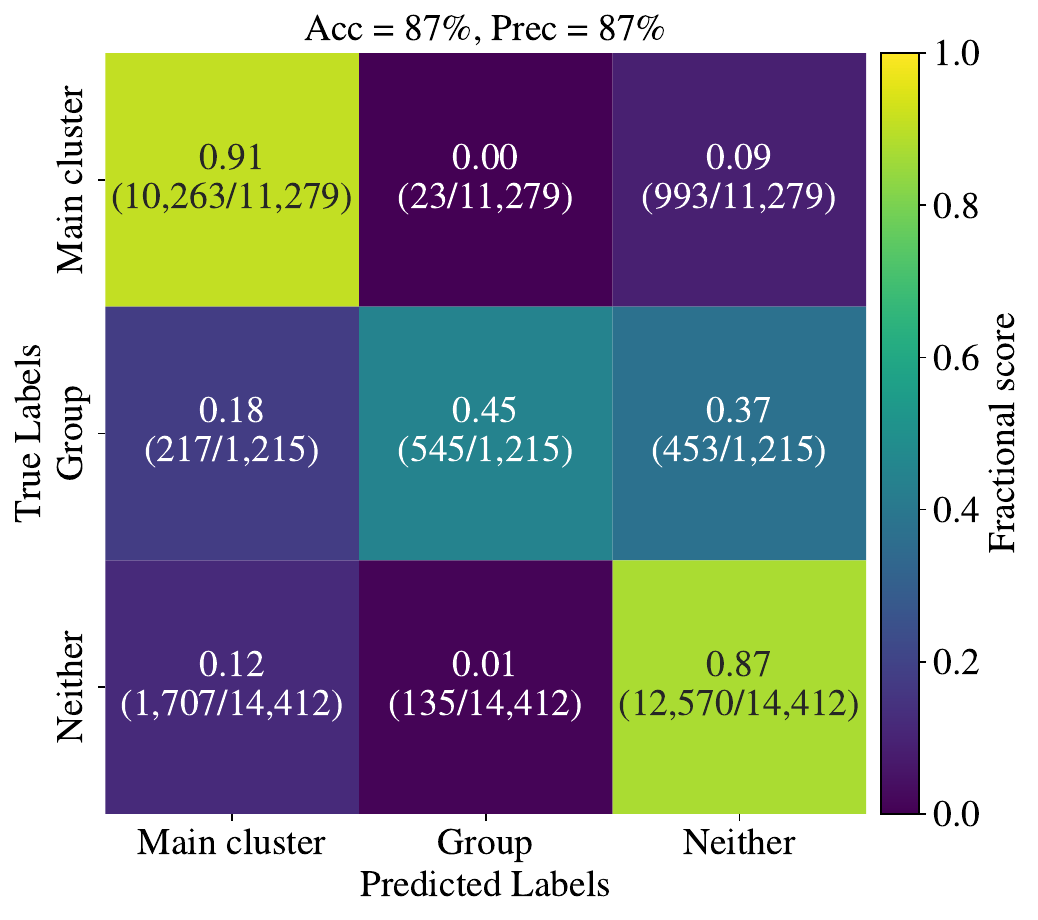}
    \caption{Confusion matrix of test set predictions from the model trained on full 3D information. Each row corresponds to galaxies from the true (labelled) class, while each column indicates the class assigned by the model. Both fractional and raw counts are displayed in each cell to convey class distributions. Notably, the \textit{group} class is under-represented, comprising roughly ten times fewer galaxies than the other classes. The reported overall accuracy reflects all predictions, and the precision values are weighted by class size.}
    \label{fig: CM 3D}
\end{figure}

We can quantify the predictions for all test clusters by calculating precision and recall scores for each class. Precision measures the reliability of positive predictions (i.e., purity), while recall quantifies the proportion of actual positives correctly identified (i.e., completeness). Predictions are classified into four categories: true positive (TP) — a target class is correctly predicted; true negative (TN) — a non-target class is correctly predicted; false positive (FP) — a non-target class is incorrectly predicted as the target; and false negative (FN) — a target class is incorrectly predicted as a non-target. These quantities are used to calculate the precision and recall as follows,
\begin{equation}
    \mathrm{Precision \;(Purity)} = \frac{\mathrm{TP}}{\mathrm{TP} + \mathrm{FP}},
\end{equation}
\begin{equation}
    \mathrm{Recall \;(Completeness)} = \frac{\mathrm{TP}}{\mathrm{TP} + \mathrm{FN}}.
\end{equation}
The total accuracy of the model can also be calculated using,
\begin{equation}
    \mathrm{Accuracy} = \frac{\mathrm{Number \;of \;correct \;predictions}}{\mathrm{Total\; number \;of \;samples}}.
\end{equation}
For all metrics, scores closer to one (100\%) indicate better performance. For consistency, we hereafter refer to precision as purity and recall as completeness, although the terms are conceptually interchangeable. Confusion matrices provide a compact summary of all classification results. The rows correspond to the true class labels, and the columns to the predicted labels. Each tile is normalised by the total number of galaxies in the true class (sum to unity row-wise). This is such that the leading diagonal indicates the completeness of each class. The off-diagonal elements represent misclassifications, which can occur as either false positives or false negatives. Figure~\ref{fig: CM 3D} shows the confusion matrix for all test cluster data.

We achieve a good overall accuracy of 87\% with a class-weighted precision of 87\% as well. We obtain strong completeness and purity scores for the \textit{neither} and \textit{main cluster} classes, with the \textit{neither} exhibiting the highest performance at 87\% completeness and 90\% purity. The \textit{main cluster} class follows closely with 91\% completeness and 84\% purity. In contrast, the \textit{group} class shows lower performance, with completeness of 45\% and purity of 78\%. Very few \textit{neither} and \textit{main cluster} galaxies are incorrectly assigned to the \textit{group} class. However, more than half of the \textit{group} galaxies are incorrectly assigned as \textit{main cluster} (18\%) and \textit{neither} (37\%). Those assigned to the \textit{main cluster} are closest to the centre in projection, whereas those assigned to the \textit{neither} class are further out and are in lower-density environments. 

The test set contains 116 distinct groups, 31 (27\%) of which have no predicted members. On average, 36\% of the galaxies in each group are recovered with a scatter of approximately 30\%. However, the model performs well in recovering group central galaxies as members of the group class, correctly classifying 77 out of 104 (72\%). The discrepancy between the number of groups and group centrals arises because some group centrals lie outside the 7.5~Mpc boundary, whereas some of their members remain within it.

\subsubsection{Optimising group galaxy predictions}\label{subsec: model optimisation}

The primary challenge in our dataset is the severe imbalance among classes, with approximate training population ratios of 12:9:1 for the \textit{neither}, \textit{main cluster}, and \textit{group} classes, respectively. Galaxy groups are capped at a maximum of 50 members, while \textit{main cluster} and \textit{neither} galaxies face no such restriction. This imbalance biases the model towards the more populous classes, compromising performance on the minority \textit{group} class, which is our main focus.

We explored several strategies to mitigate this, including increasing the training weight of the \textit{group} class, resampling (via bootstrap oversampling and random undersampling), and adopting a two-step modelling approach to first separate low- and high-density environments before identifying \textit{group} galaxies. However, none of these approaches led to meaningful improvements (see section \ref{sec: 2D observations} for further discussion). The fundamental issue is that \textit{group} galaxies are not sufficiently separable in feature space; strong overlap with other classes limits the effectiveness of weighting and resampling strategies (see Figures~\ref{fig: 3D sim feature space distns} and \ref{fig: 2D obs feature space distns} in appendix~\ref{sec: appendix A} for the extent of the overlap).

We also examined adjusting the probability thresholds for assigning galaxies to the \textit{group} class, optimising for different metrics such as purity and completeness using variations of the F$_\beta$-score. This is harmonic mean of the purity and completeness and is calculated using,
\begin{equation}
    \mathrm{F}_{\beta} = \frac{(1+\beta^2)\mathrm{TP}}{(1+\beta^2)\mathrm{TP} + \mathrm{FP} +{\beta^2}\mathrm{FN}},
\end{equation}
where $\beta$ parametrises the relative importance between purity and completeness. A $\beta$ value of 1 represents equal purity and completeness, where $\beta \rightarrow 0$ increasingly prioritises purity, and $\beta \rightarrow \infty$ towards completeness. In practice, for our dataset, adjusting the classification threshold trades completeness for purity (and vice versa) in an almost linear fashion. Consequently, there is no meaningful compromise to be achieved between these two metrics, particularly given the underlying class entanglement. Appendix~\ref{sec: appendix B} goes into more detail on optimisation methods given the validation cluster results.

In all cases, the default first-past-the-post approach, assigning each galaxy to the class with the highest predicted probability, combined with training on the original imbalanced dataset, ultimately performs best for our scientific objectives. This strategy yields high purity in \textit{group} predictions, which we prioritise to ensure reliable class membership. However, in scenarios where higher completeness is desired, such as selecting more targets for follow-up observations, a different balance may be preferable. Our method is flexible, and it would be straightforward to re-prioritise completeness over purity if the scientific application requires it.

Nevertheless, for the remainder of this study, we adopt the model trained on the original class-imbalanced data with default probability assignments and no post-processing optimisation. This choice offers a robust balance that aligns with our emphasis on purity.

\subsection{2D Observations}\label{sec: 2D observations}

\begin{figure}
    \centering
    \includegraphics[width=\linewidth]{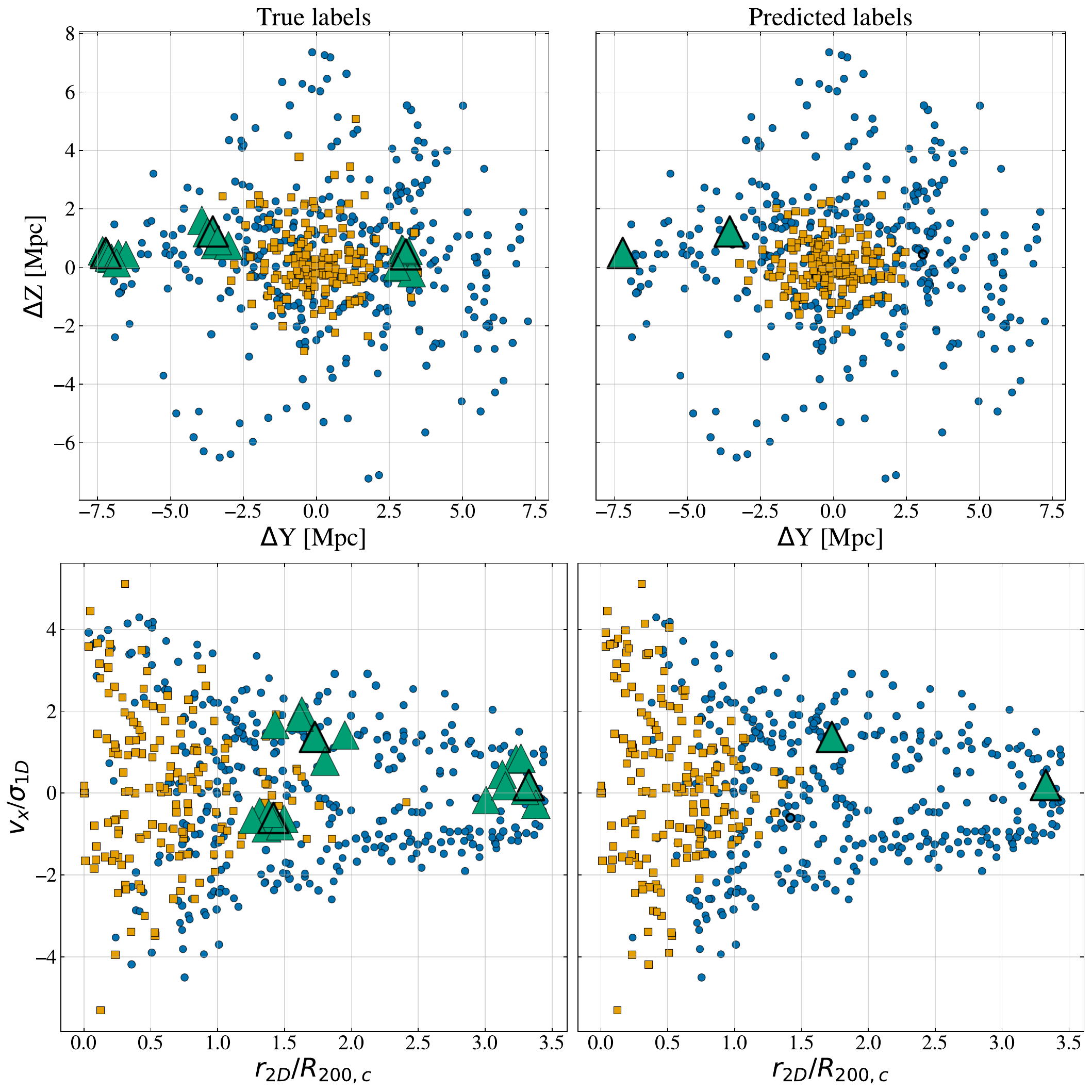}
    \caption{The same cluster and projection as Figure~\ref{fig: cluster 155 yz plane 3d}, this time trained and predicted on projected clusters. The velocity axis in the phase-space row represents the simple LoS velocity with respect to the cluster, normalised to the model 1D cluster velocity dispersion.}
    \label{fig: cluster 155 yz plane 2d}
\end{figure}
Figure~\ref{fig: cluster 155 yz plane 2d} shows the model predictions for the same cluster presented in Figure~\ref{fig: cluster 155 yz plane 3d}, now with a feature space derived from its projected counterpart, using sky positions and LoS velocities for the galaxies. The radial boundary remains fixed at 7.5~Mpc, but the projection introduces more interloper galaxies, which will skew feature space values. There is also an extra group within the projection, with only the two left hand group central being identified to the group class. All other group galaxies have been lost to the other two classes.

\begin{figure}
    \centering
    \includegraphics[width=1\linewidth]{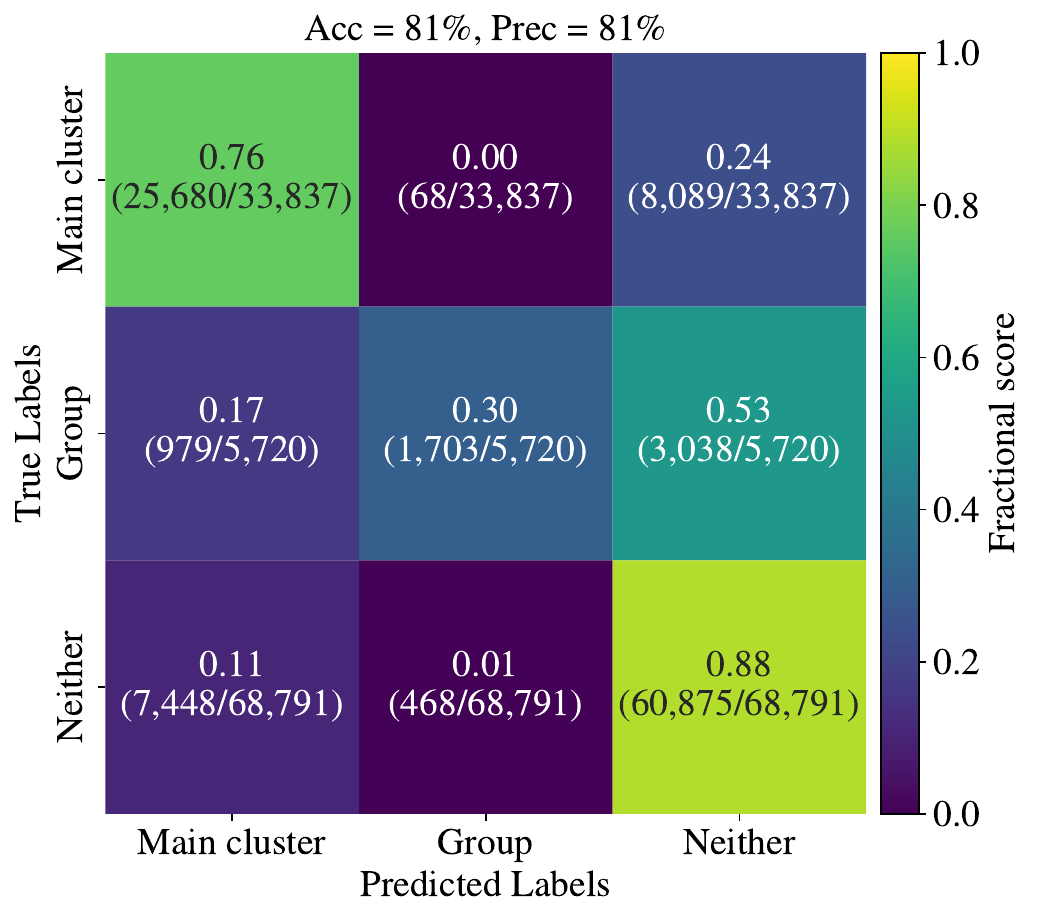}
    \caption{Confusion matrix of test data predictions using a model trained on projected 2D data with unbalanced class sizes and first-past-the-post class assignments.}
    \label{fig: CM 2D}
\end{figure}

Figure~\ref{fig: CM 2D} shows the confusion matrix for all test clusters. As expected, applying the model to observational data results in a drop in overall performance, with accuracy reduced to 81\% and weighted precision to 81\%. Class-specific metrics also decline. The model remains most effective at identifying \textit{neither} class galaxies, achieving 88\% completeness and 85\% purity. Performance on the \textit{main cluster} class follows, with 76\% completeness and 75\% purity. The \textit{group} class remains the most challenging, with completeness at 30\% and purity at 76\% for groups at all cluster-centric radii (see section 4.5 for radial dependence). This result corresponds to the maximum F$_{\beta}$ score at an effective $\beta$ of approximately 0.34, indicating that purity (precision) is implicitly valued about three times more than completeness (recall). The effective $\beta$ thus summarises the trade-off point on the PR curve nearest to the fixed decision threshold, highlighting which F$_{\beta}$ score would naturally favour this point as optimal.

\begin{figure*}
    \centering
    \includegraphics[width=0.9\linewidth]{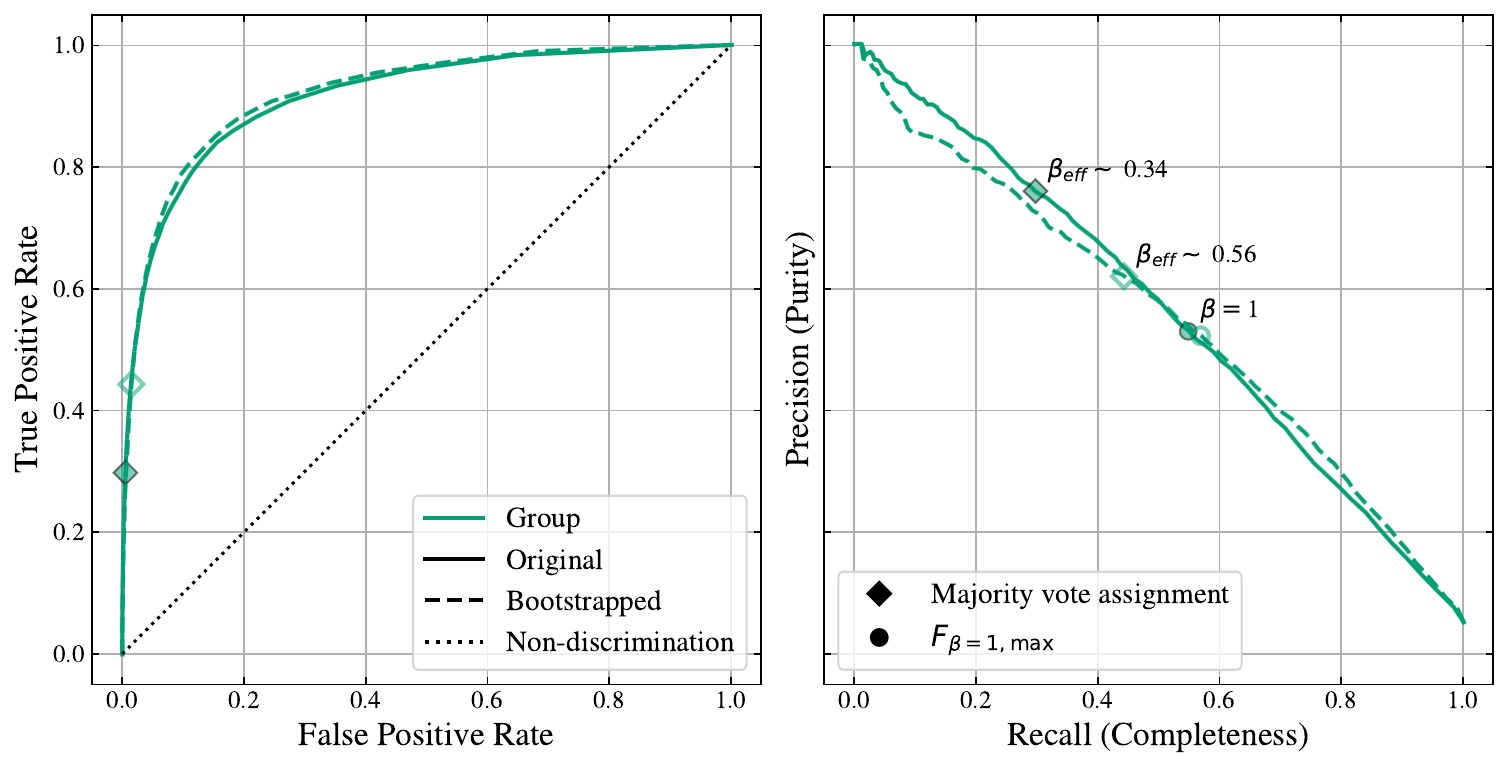}
    \caption{ROC (\textit{left}) and PR (\textit{right}) curves for the \textit{group} class, evaluated using the RF model trained on 2D projected clusters. The original class-imbalanced case is shown with solid lines and filled markers, while the bootstrap class-rebalanced model is shown with dashed lines and hollow markers; both are tested on the corresponding test clusters. For a reference of model performance on the \textit{main cluster} and \textit{neither} classes, see Figure~\ref{fig: 3D sim ROC and PR plots} (these are for the 3D datasets, though the trends are similar). The black dotted line in the ROC plot represents the performance of a classifier no better than random guessing. Diamond markers denote the points corresponding to the default highest-probability (majority vote) assignment. Additional markers indicate the optimal points for each associated performance metric, providing direct visual references for comparing model behaviours. The annotated $\beta$ values in the PR plot specify the hypothetical effective $\beta$ at which the F$_\beta$ score is maximised, highlighting the implicit precision–recall trade-off at each point.}
    \label{fig: 2D test clusters ROC and PR curves}
\end{figure*}

To emphasise the challenges of model optimisation and the trade-off between purity and completeness discussed in section~\ref{subsec: model optimisation}, we present the receiver operating characteristic (ROC) and precision–recall (PR) curves for the test cluster predictions in Figure~\ref{fig: 2D test clusters ROC and PR curves}.

The ROC curve illustrates the trade-off between the true positive rate (sensitivity) and the false positive rate, indicating how well the model identifies the group class. An ideal classifier achieves an area under the curve (AUC) of 1 and passes through the point (0, 1). The PR curve shows the relationship between purity (precision) and completeness (recall) across various assignment thresholds. Diamond markers denote the default first-past-the-post prediction points for reference. For an ideal classifier, the PR curve also has an AUC of 1 and passes through the point (1, 1).

In our results, the curves for both class-size sampling strategies described in section~\ref{subsec: model optimisation} closely overlap, indicating minimal differences in overall discrimination power. In the ROC space, both default assignment points lie well below the ideal point, highlighting suboptimal performance under this metric for group galaxy selection (the global performance when considering all galaxy classes is much better, as shown in Appendix~\ref{sec: appendix B}). In the PR space, these points lie in the high-purity region. The cost of group completeness is primarily attributed to the class imbalance during training, which is reflected in the low effective $\beta$ score. Notably, at the balanced point where purity and completeness are equally weighted ($\beta=1$), both models perform almost identically. Given the negligible differences and our scientific preference for purity, we retain the original class-imbalanced model instead of using the class-balanced alternatives.

The observational test set contains 498 groups, with 145 (29\%) having no predicted members. The mean group completeness is 25\%, with a standard deviation of 23\%. As for the group central galaxies, 337 out of 460 (73\%) have been correctly identified as part of the group class. Overall, the model performs worse when trained on realistic 2D projected clusters, so training on the full 3D model is much preferred.

\begin{figure}
    \centering
    \includegraphics[width=1\linewidth]{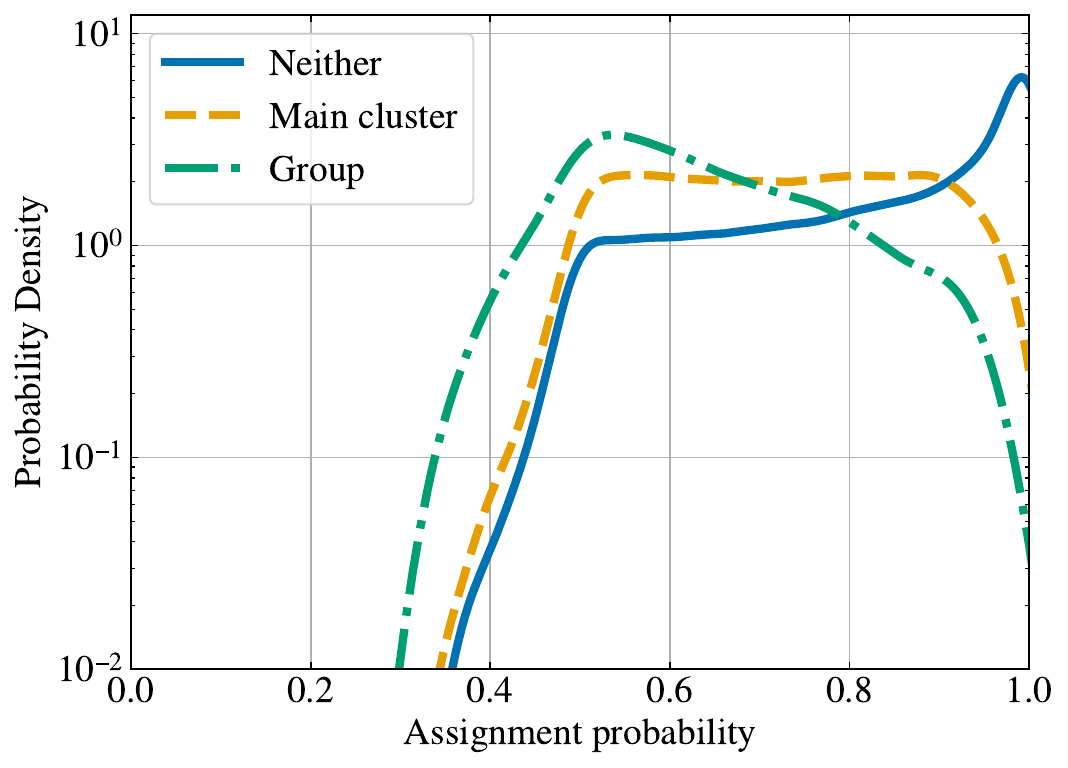}
    \caption{Smoothed probability distributions of galaxies assigned to each class in observations. The blue solid line corresponds to galaxies assigned to the \textit{neither} class, the orange dashed line to the main cluster, and the green dash-dotted line to the group.}
    \label{fig: 2D obs PDFs}
\end{figure}

In Figure~\ref{fig: 2D obs PDFs} we show the probability density functions (PDFs) of the galaxies' class assignment probabilities extracted from the RF classifier and smoothed with a Gaussian kernel. The PDF for the \textit{neither} class shows a sharp peak beyond 90\%, reflecting high model confidence. Combined with strong completeness and purity scores, this indicates the model reliably identifies unbound galaxies. The \textit{main cluster} predictions are flat between 50\% and 90\%, followed by a steep decline on either side of the plateau. Nonetheless, the high classification metric values again suggest the model successfully recovers galaxies gravitationally bound to the main cluster. In contrast, the \textit{group} class PDF peaks near 50\% and falls off sharply at higher probabilities, where the remaining high-probability tail predominantly corresponds to the prediction of the group hosts (central group galaxies).

\subsection{Feature Importances}

\begin{figure}
    \centering
    \includegraphics[width=1\linewidth]{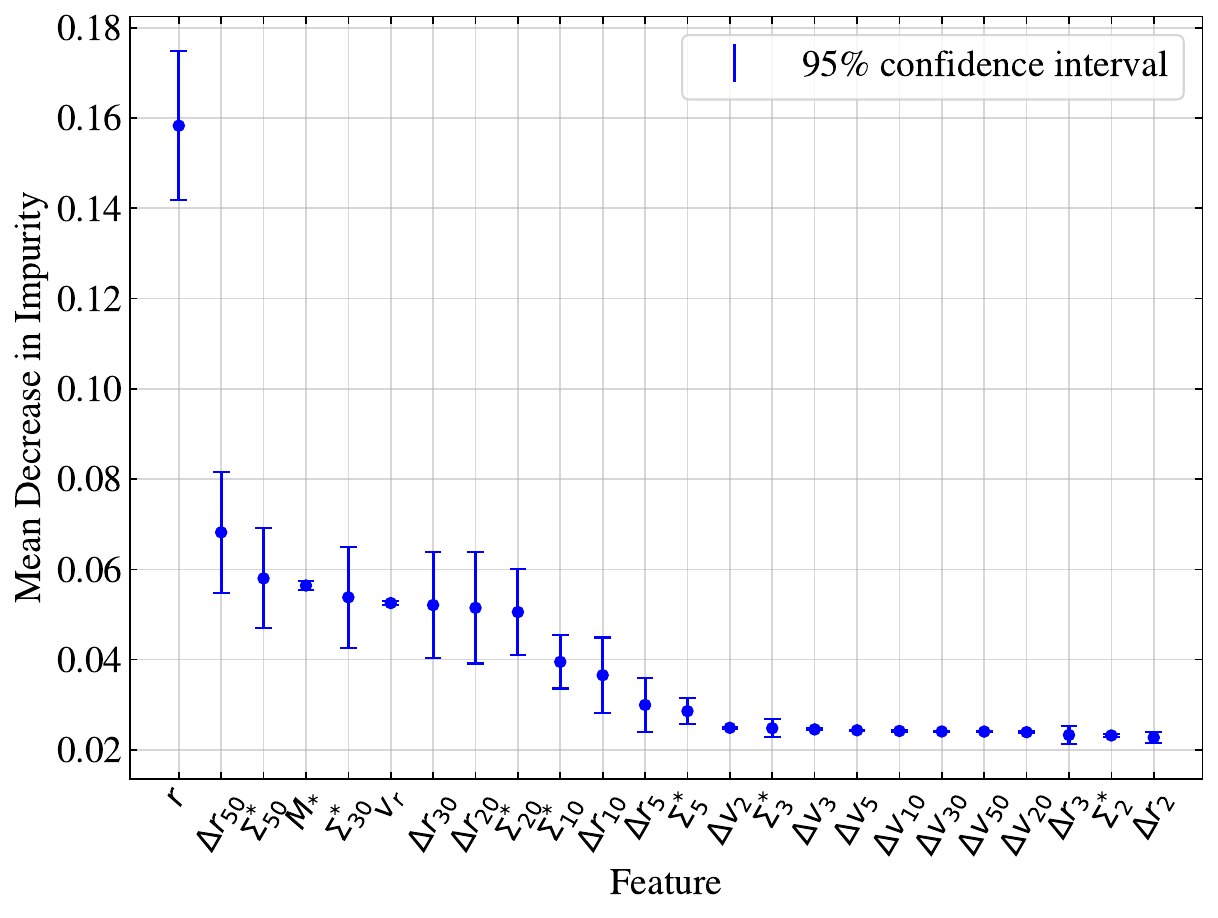}
    \caption{Feature importance for all observational features in descending order. Data points represent the final MDI across all trees, with errors representing the 95\% confidence interval.}
    \label{fig: 2D obs feature importances}
\end{figure}

During training, splits are made within each feature-space input to maximise class purity. Once the best split is made, the impurity of the remaining class data will decrease. The mean decrease in impurity (MDI) for a given feature across all models in the ensemble quantifies how `important', i.e., useful, the feature is to model training. Higher values reflect more useful features. This helps us to understand what the model learns during training. Figure~\ref{fig: 2D obs feature importances} shows the rank-ordered relative MDI for each feature within our 2D observational feature space (see Section~\ref{sec: 2D obs FS} for feature space generation).

The cluster-centric radial distance of a galaxy is the most useful feature of all. This is because of the prominent divide between the \textit{main cluster} and \textit{neither} galaxy radial positions (see Figure~\ref{fig: cluster 155 yz plane 2d} for an example). By definition, \textit{main cluster} galaxies reside close to the cluster centre (typically, the Brightest Cluster galaxy, BCG) in all projections. As a result, a radial split in the data (e.g., at about 2~Mpc in plot A of Figure~\ref{fig: 2D obs feature space distns}), would produce a relatively pure separation of \textit{main cluster} galaxies; therefore, a significant decrease in data set impurity. The subsequent useful parameters are the high-order local densities, meaning the model generally captures the presence of over- and under-dense regions well. High-order galaxy separation is beneficial, as it also provides insight into the local environment. Though not as sensitive as mass density, the target stellar mass is highly rated as it contributes to the projected local density and correlates with over-dense structures. Note that, since galaxy stellar mass strongly correlates with the galaxy environment \citep{Balogh:2001, Baldry:2006, Etherington:2017}, quantifying the environment using the galaxy's own mass is somewhat circular. Section~\ref{sec: removing-target-mass} will explore this further. The cluster-centric radial velocity is helpful due to the collective differences between infalling, non-infalling, and orbiting (virialised) galaxies. On large scales, field galaxies have coherent velocities comparable to the Hubble flow, which, when centred on the cluster, will be apparent to the model (seen in plot I of figure~\ref{fig: 2D obs feature space distns}). Infalling galaxies, e.g., via filamentary networks, will have a boosted coherent velocity within the vicinity of the cluster. The main issue, especially with galaxy-galaxy relative velocities, is signal washout due to the peculiar motions of orbiting galaxies in more virialised systems, such as clusters and groups. The Fingers-of-God effect will obscure coherent motion, rendering the velocity information relatively ineffective in observations. This is why the remaining features have significantly lower importance. This interpretation is supported by a direct test: when retraining the model using only the ten most important features, we observe only a marginal decrease in performance. This indicates that the excluded features -- particularly the velocity-related ones -- contribute little to the model’s predictive power.

\subsection{Removing target galaxy mass}\label{sec: removing-target-mass}

\begin{figure}
    \centering
    \includegraphics[width=1\linewidth]{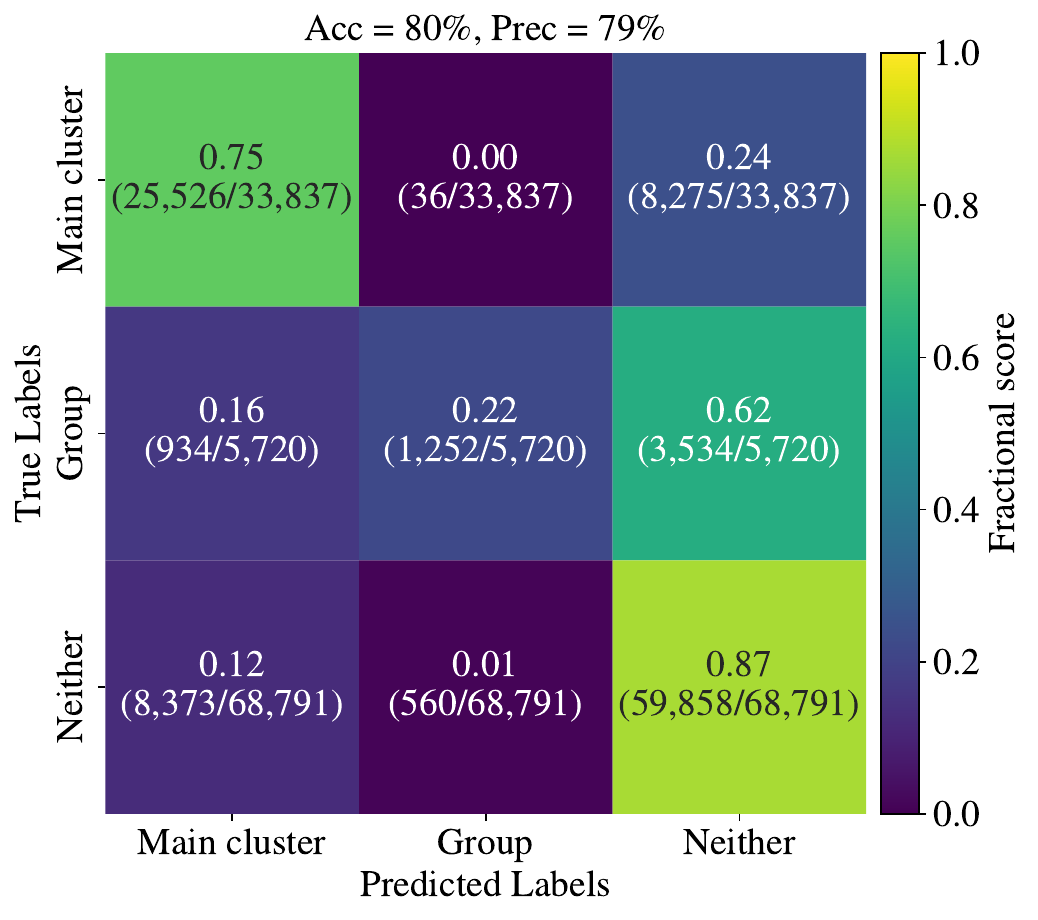}
    \caption{Confusion matrix for 2D projected test clusters with target stellar mass removed from feature space.}
    \label{fig: 2D obs CM no target mass}
\end{figure}

As mentioned in the previous section, if one's science objective is to investigate intrinsic galaxy properties within different environments, then stellar mass should not be used to define that environment. In this case, we recommend removing target stellar mass as an individual feature, and from local density calculations, i.e., start the summations at $i=2$ in Equations~\ref {eqn: mass-weighted local density 3D} and \ref{eqn: mass-weighted local density 2D}. Figure~\ref{fig: 2D obs CM no target mass} shows the updated confusion matrix reflecting this change.

Removing the target stellar mass feature results in a negligible decline of 1\% in overall accuracy and 2\% in weighted precision. More notably, \textit{group}-class completeness and purity decrease by 8\% each, while \textit{neither}-class predictions fall by 1\% in both completeness and purity. \textit{Main cluster} predictions show a 1\% drop in completeness and a 2\% reduction in purity.

Target stellar mass is informative on its own; however, its removal does not substantially degrade overall model performance because its predictive power is partially compensated by high-order local density features. For instance, excluding a single galaxy from a 10-nearest-neighbour density estimate typically alters the result by about 10\% on average. This sensitivity decreases for higher-order density estimates, making the model generally robust to such perturbations. Nonetheless, the \textit{group} class is most adversely affected by the exclusion of stellar mass. Consequently, omitting stellar mass has a non-negligible impact on \textit{group} predictions and should be carefully considered when the aim is to isolate environmental effects from intrinsic galaxy properties.

\subsection{Model performance vs cluster and galaxy properties}

\begin{figure}
    \centering
    \includegraphics[width=1\linewidth]{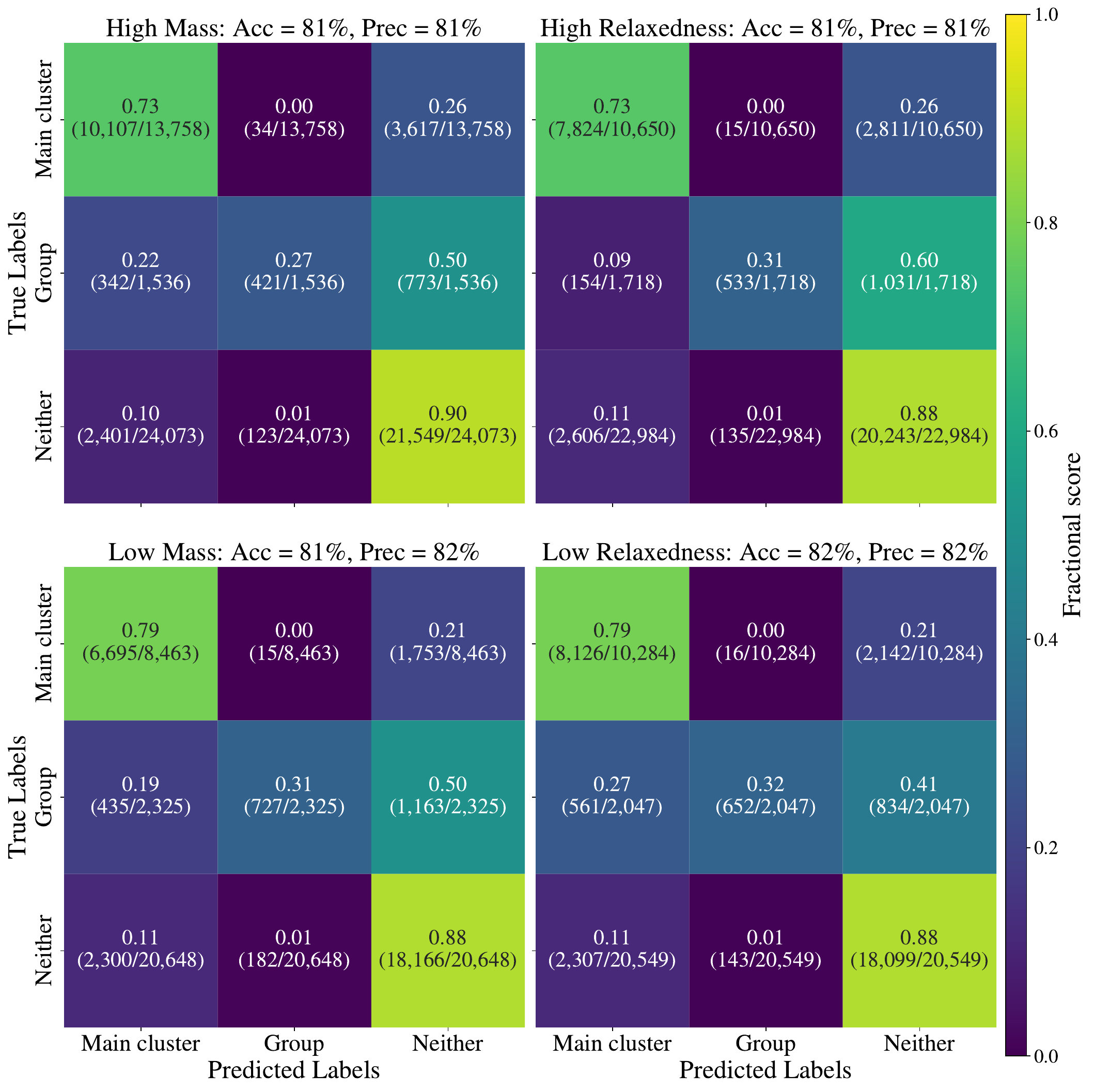}
    \caption{Confusion matrices for high mass (\textit{top left}), high relaxedness (\textit{top right}), low mass (\textit{bottom left}) and low relaxedness (\textit{bottom right}. As for the previous matrices, the rows correspond to true labels, and the columns show class predictions.}
    \label{fig: 2D obs extrema CMs}
\end{figure}

To be practically useful, machine learning models must generalise well, minimising the need for extensive pre-processing or retraining when applied to new datasets. To evaluate generalisability, we trained 10 separately initialised models on the same data and found that the variation in predictions was minimal, with scatter of the order of 1\% in the different performance metrics.

We further tested model robustness by applying the first trained model to clusters in the upper and lower thirds of the mass and dynamical relaxation distributions. Confusion matrices for these four extreme cases are shown in Figure~\ref{fig: 2D obs extrema CMs}. Model performance remained consistent across these subsets, which is unsurprising given that the training, validation, and test sets exhibited statistically compatible mass and relaxedness distributions ($p$ is always above 0.015 in the KS test).

\begin{figure*}
    \centering
    \includegraphics[width=1\linewidth]{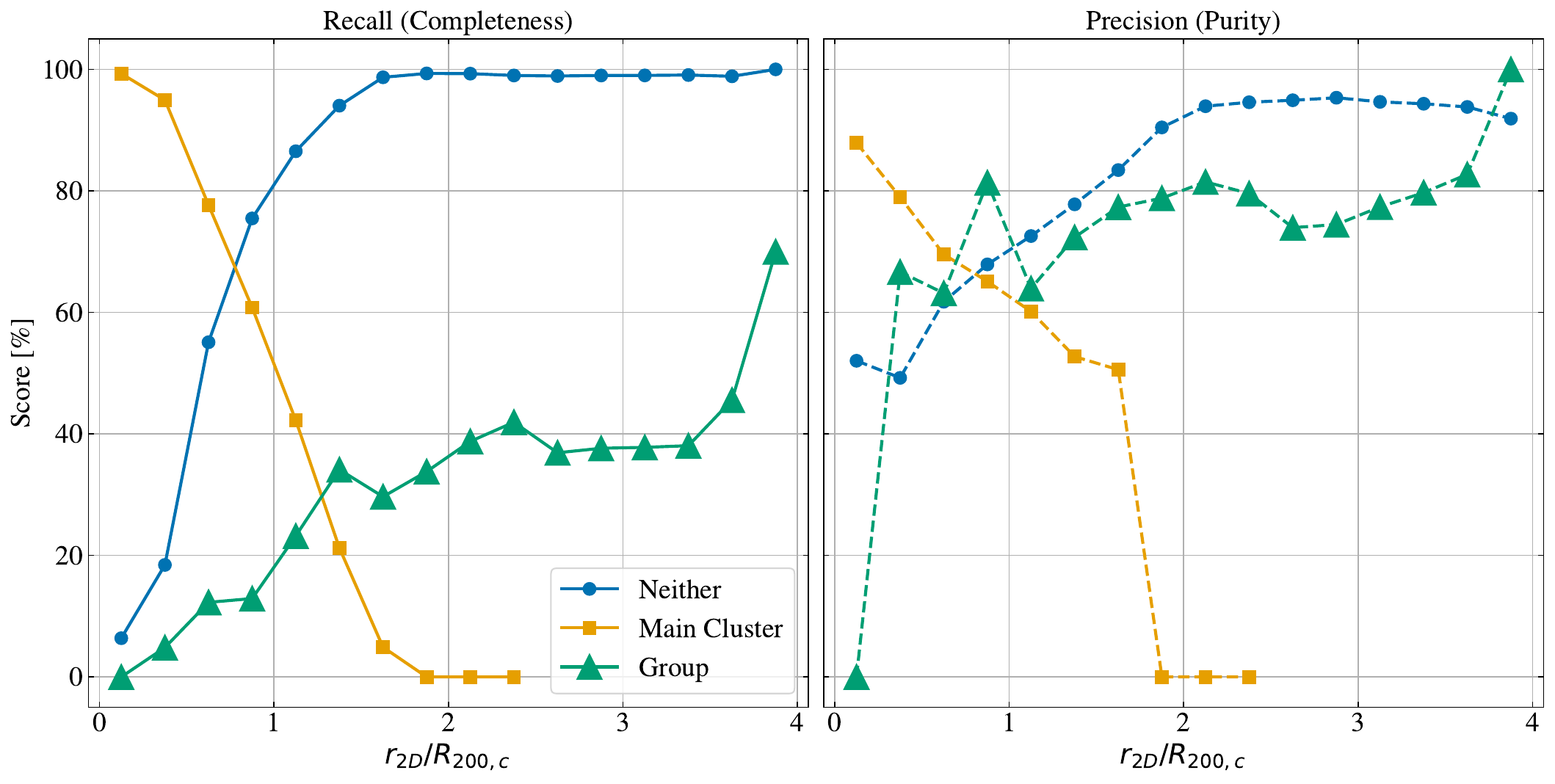}
    \caption{Model, trained on 2D projected clusters, completeness (\textit{left}: solid) and purity (\textit{right}; dashed) as a function of cluster-centric radial distance, normalised to respective cluster $R_{200,c}$. The blue circles represent \textit{neither} class performance; the orange squares are for \textit{main cluster}, and the green triangles are for \textit{group}.}
    \label{fig: 2D obs R and P vs r}
\end{figure*}

Since our objective is to identify merging and infalling groups, we assessed the model performance as a function of galaxy radial distance to the cluster core, with the completeness and purity curves shown in Figure~\ref{fig: 2D obs R and P vs r}.

Within clusters, model performance varies significantly with galaxy radial distance from the cluster centre, across all classes. As expected, \textit{main cluster} galaxies exhibit the highest completeness and purity near the core, with both metrics declining steeply beyond the centre. The \textit{main cluster} classification curves truncate shortly after $2\times R_{200,\mathrm{c}}$, due to the imposed backsplash limit at $2.5 \times R_{200}$ (see description of catalogue generation in Section~\ref{subsec: substructure catalogues}), beyond which no galaxies are considered bound to the main halo.

In contrast, \textit{neither} galaxy prediction reaches near-perfect recovery beyond $2\times R_{200,\mathrm{c}}$, but declines sharply at smaller radii, reversing the trend seen for \textit{main cluster} galaxies. Group galaxy prediction completeness is around 40\% and purity 70\% within the infall region ($1.5 \lesssim r\,[R_{200,c}] \lesssim 3.5$), with both metrics steadily increasing at larger radii. As a result, our model will be biased towards identifying groups in the outskirts. Beyond $\simeq 4 \times R_{200}$, group completeness and purity increase very significantly: finding group galaxies far away from massive clusters is much easier, and our method, together with simpler alternatives such as friends-of-friends, works well. Within the cluster is where small to intermediate-sized groups become disrupted and lose most of their members to the cluster \cite{Haggar:2023}; finding groups here is shown to be challenging.

To further characterise potential selection effects, we examine model performance as a function of galaxy stellar mass by comparing the stellar mass distribution of the ``true'' group galaxy population with that recovered by our method. We find that the recovered sample is biased towards higher stellar masses. Dividing galaxies into three bins: low ($M_\ast < 10^{10} M_\odot$), intermediate ($10^{10} M_\odot \leq M_\ast < 10^{11} M_\odot$), and high ($M_\ast \geq 10^{11} M_\odot$), yields completeness and purity values of $(0.18, 0.69)$, $(0.33, 0.74)$, and $(0.66, 0.87)$, respectively. This demonstrates a clear systematic bias, with substantially improved recovery of high-mass group galaxies compared to their lower-mass counterparts. We will address this bias and practical ways to alleviate it in a subsequent paper (Jordan et al. 2026, \textit{in prep}.).

\subsection{Comparison to common methods}

To motivate our choice of supervised machine learning for identifying group galaxies in the infall region, we compared its performance against two widely used substructure-finding techniques: Friends-of-Friends (FoF) and Gaussian Mixture Modelling (GMM). The FoF algorithm links galaxies separated by less than a distance defined relative to the large-scale background density. This works well for detecting overdensities in redshift surveys, where the background density is approximately uniform and low. A common choice is to set the perpendicular linking length to $0.2$ times the mean inter-galaxy separation. However, when applied to the infall-region field of view around each cluster, FoF produces very low group completeness and purity, typically only $5$--$10\%$. The poor performance stems from the highly non-uniform background density in these regions, which makes the standard linking length inappropriate. Adding an additional radial linking length exacerbates the problem, as redshift-space distortions stretch galaxy distributions along the line of sight.

In contrast, GMM models the combined sky position and line-of-sight velocity distributions of galaxies using a mixture of Gaussian probability density functions, referred to as ``components.'' The number of components is varied, with the optimal model chosen according to the Bayesian Information Criterion \citep{Kass:1995}. Each Gaussian component corresponds to a candidate substructure; however, by construction, all galaxies are assigned to a component, fixing the nominal completeness at 100\%. To address this, we apply probability cuts to refine galaxy membership, retaining only those most likely to belong to significant dynamical structures within the cluster region. An example of an optimal model fit to a cluster projection, together with the corresponding analysis of completeness and purity as a function of probability cut, is presented in Appendix~\ref{sec: appendix C}. The GMM method yields inherently high group completeness, but this comes at the expense of very low group purity, driven by extreme contamination from non-group galaxies, even under stringent probability thresholds for Gaussian components. Consequently, GMM is unsuitable for our scientific objectives.

\section{Conclusions}\label{sec: conclusions}

The environment in which a galaxy resides plays a pivotal role in shaping its evolutionary path. Interactions with the intragroup or intracluster medium, such as ram pressure, can significantly alter galaxy properties. Proximity to deep gravitational potential wells, such as those found in galaxy clusters, further intensifies these environmental effects. Clusters encompass many different environmental conditions, from the relatively sparse outer field regions to the dense, dynamic cores. To investigate the diverse physical processes at play, it is essential to identify a galaxy’s environment reliably in observational data. However, this task is challenging within the infall region and cluster ($<4\times R_{200}$) due to projection effects in both spatial and redshift dimensions, notably the FoG phenomenon. To address this, we employed cosmological zoom resimulations of massive galaxy clusters to train a supervised machine learning model capable of classifying galaxies into three environmental categories: \textit{main cluster}, \textit{group}, and \textit{neither}. The model was optimised to maximise \textit{group} class purity, as this yields more reliable \textit{group} galaxy predictions. This is particularly desirable when comparing the properties of group galaxies with those of galaxies in other environments: clean samples are more important than complete ones, particularly when the samples are large. However, the choice of prioritising purity over completeness ultimately depends on the specific scientific objective. For example, when selecting targets for follow-up observations to confirm group membership, a more pragmatic balance between completeness and purity may be preferable. Our method is flexible enough to allow implementing such a choice if required.

We also evaluated the model’s performance separately within the cluster core and the surrounding infall regions, which are challenging yet key environments. Our random forest model takes projected galaxy phase space and stellar mass information as inputs to learn mappings and make class predictions. We summarise our main findings below.
\begin{enumerate}
    \item Our model, trained on 2D projected clusters using a feature space derived from galaxy properties that can be derived from observations (positions on the sky, line-of-sight velocities, and stellar masses), achieves an overall accuracy and class-size-weighted precision of 81\%. It performs best at classifying galaxies into the \textit{neither} class, representing unbound galaxies. Membership of the \textit{main cluster} is also well recovered, with 76\% completeness and 75\% purity. However, the model struggles with \textit{group} galaxy predictions, achieving only 30\% completeness despite a relatively high purity of 76\%. The resulting model has an effective $\beta$ of 0.34 (cf. section~\ref{sec: 2D observations}), implying a preference for purity that is approximately three times stronger than for completeness, a trade-off that aligns with our scientific objectives. The model misses all members from 145 of 498 (29\%) of the groups, but disproportionately recovers 337 of 460 (73\%) group hosts (central group galaxies). This is promising and can act as a seed for the follow-up study by identifying individual groups (similar to the methodology in \citealt{Ma:2025}). 

    \item Resampling strategies that address the imbalance between the galaxy class sizes typically improve completeness for under-represented classes such as group galaxies, but this comes at an almost linear cost to purity, with the trade-off worsening as the initial class size decreases. Adjusting training weights in favour of minority classes offered little benefit for smaller populations due to substantial feature overlap between classes. This overlap is primarily driven by projection effects, which distort both local density estimates and velocity measurements. However, even in 3D simulation, there is an intrinsic similarity between \textit{main cluster} and \textit{group} galaxy neighbourhood properties.

    \item The most informative feature was radial distance to the cluster centre, which cleanly separates \textit{main cluster} and \textit{neither} galaxies. Next in importance were local mass-weighted densities and sky separations using ten or more nearest neighbours, while lower-order neighbour features contributed little. Among velocity features, radial velocity relative to the cluster was most valuable, reflecting coherent infall from gravitational collapse. These velocity-based features were especially effective at distinguishing \textit{neither} galaxies from the \textit{main cluster} and \textit{groups}.

    \item The model exhibits no significant bias with respect to cluster mass or dynamical state but shows a clear radial dependence in classification performance. As expected, \textit{main cluster} galaxies are most reliably identified near the cluster core. \textit{neither}-type galaxies are best recovered at large radii, with near-perfect completeness in the infall region. Group galaxy predictions follow a similar radial trend to \textit{neither}-type galaxies but with consistently lower performance. Within the cluster ($\leq 1 \times R_{200,c}$), the model identifies roughly 15\% of \textit{group} galaxies at 60\% purity. Beyond the cluster boundary, completeness rises to ~40\%, while purity stabilises around 80\%. This radial dependence is robust, with low scatter across ten independently trained models.
\end{enumerate}

This study demonstrates that machine learning provides a powerful and flexible framework for addressing complex, data-driven problems, such as identifying substructures, such as galaxy groups, in observational data using cosmological simulations. While observational constraints—particularly near the cluster core and in the infall region—limit classification performance, we show that high class purity is achievable, which is essential for the reliable identification of \textit{group} galaxies. Although completeness is relatively low in the complex infall region, identifying very clean samples of group galaxies will allow for a reliable comparison of their properties with those of galaxies in other environments. With the large samples new spectroscopic surveys will provide, high purity is more important than completeness.

While our results are derived entirely from simulations, applying such models to real observational data presents additional challenges. However, previous work has shown that this gap can be mitigated with carefully designed survey strategies that adequately sample the galaxy distribution in and around clusters. For example, \citet{Cornwell:2022} demonstrated that cosmic structure models based on simulations retain strong performance when applied to realistic mock observations constructed to replicate surveys such as the WEAVE Wide Field Cluster Survey \citep{Jin:2024}.

Our approach is therefore well-suited to upcoming spectroscopic surveys, which will offer critical insight into the processes governing galaxy evolution across a range of cluster environments.

\section*{Acknowledgements}

This work was supported by the School of Physics and Astronomy at the University of Nottingham.

We acknowledge access to the theoretically modelled galaxy cluster data via The Three Hundred (\href{https://the300-project.org}{https://the300-project.org}) collaboration. The simulations used in this paper have been performed in the MareNostrum Supercomputer at the Barcelona Super-computing Center, thanks to CPU time granted by the Red Espa\^nola de Supercomputaci\'{o}n. As part of The Three Hundred project, this work has received financial support from the European Union’s Horizon 2020 Research and Innovation programme under the Marie Skłodowska-Curie grant agreement number 734374, the LACEGAL project. 

We would like to extend our thanks to the anonymous referee, whose insightful comments helped improve this paper.

For the purpose of open access, the authors have applied a Creative Commons attribution (CC BY) licence to any Author Accepted Manuscript version arising.

The authors contributed as follows: RJ, MEG, and AAS formed the core team. RH and FRP provided data products as well as technical advice on the data. SPB provided expert advice on the machine learning methodology. RJ analysed the data, produced the plots, and wrote the paper with contributions from MEG, AAS, SPB, RH, and FRP.

\section*{Data Availability}

Data available on request to \textsc{TheThreeHundred} collaboration:
\href{https://www.nottingham.ac.uk/astronomy/The300/index.php}{https://www.nottingham.ac.uk/astronomy/The300/index.php}


\bibliographystyle{mnras}
\bibliography{bibliography}



\appendix

\section{Feature space distributions}\label{sec: appendix A}

\begin{figure*}
    \centering
    \includegraphics[height=0.96\textheight]{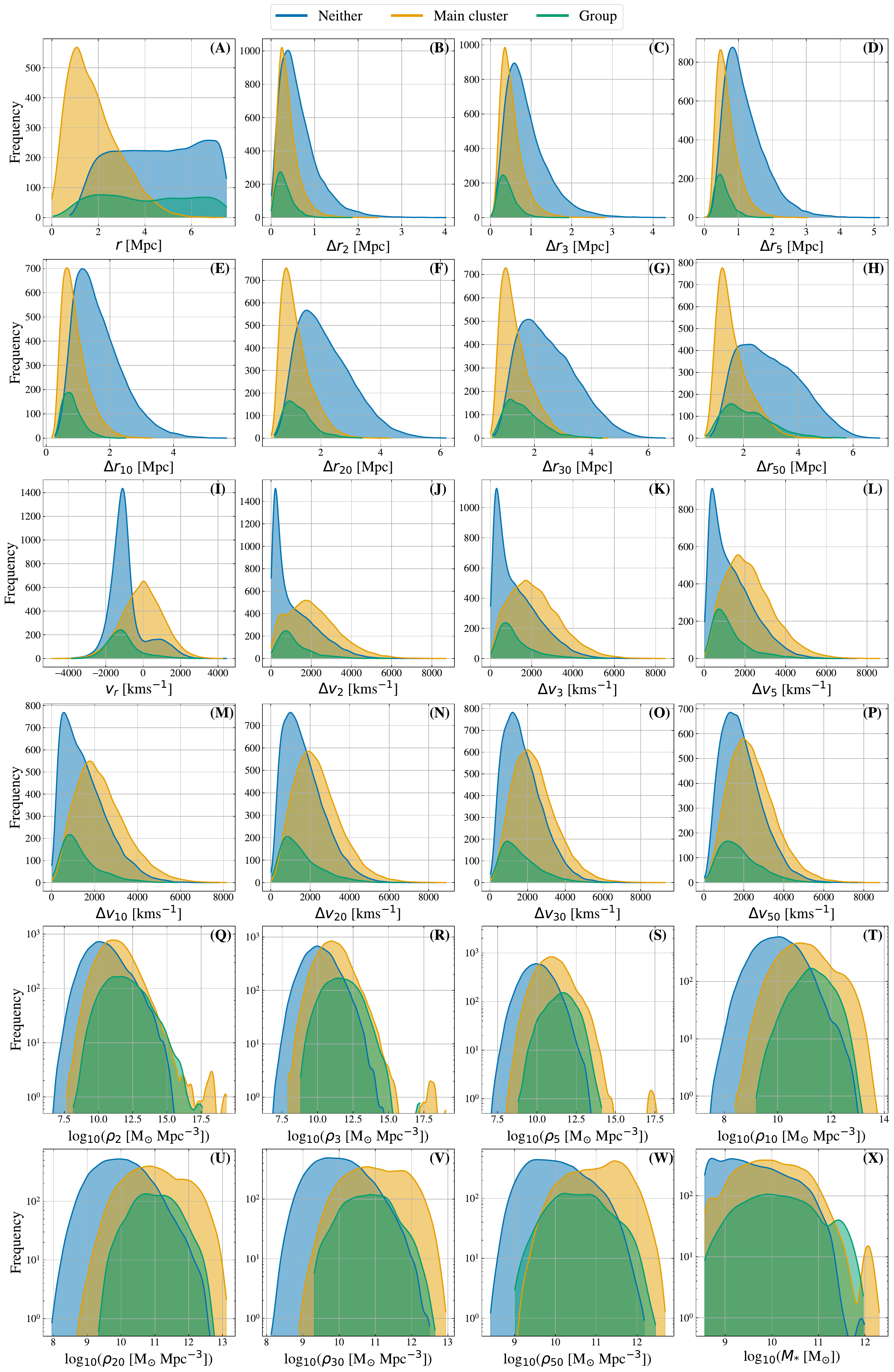}
    \caption{Distributions for each feature broken down by class in the 3D simulation training cluster feature-space.}
    \label{fig: 3D sim feature space distns}
\end{figure*}

\begin{figure*}
    \centering
    \includegraphics[height=0.96\textheight]{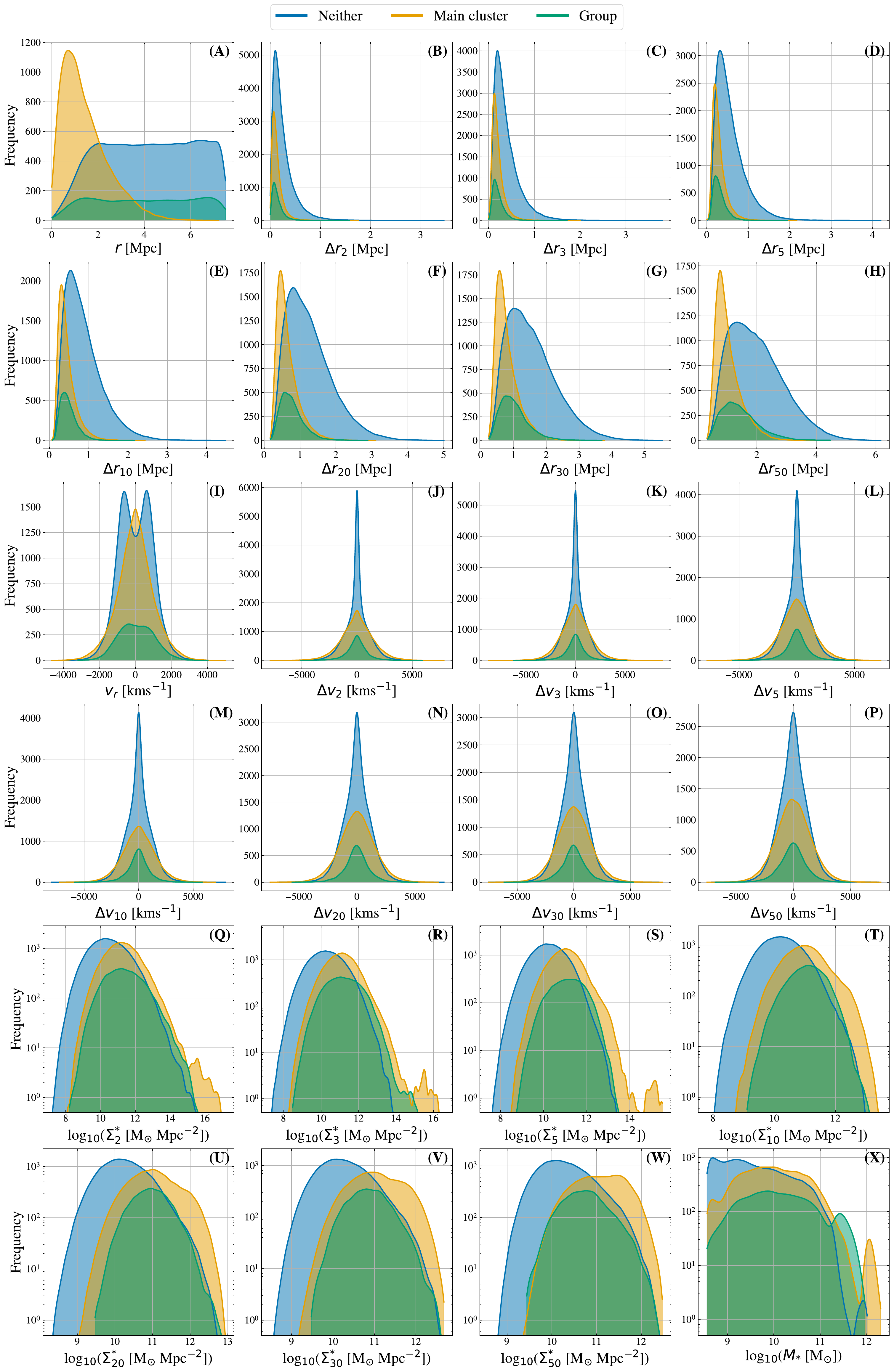}
    \caption{Distributions for each feature broken down by class in the 2D projection training cluster feature-space.}
    \label{fig: 2D obs feature space distns}
\end{figure*}

Presented in Figures~\ref{fig: 3D sim feature space distns} and \ref{fig: 2D obs feature space distns} are the class-specific feature distributions for each feature space. These highlight where there is significant overlap between the classes, hence why some features are more useful for model training than others.

We note in passing that in our initial exploration of different galaxy classification methods, we applied Principal Component Analysis (PCA) to the full 2D feature space (shown in Fig.~\ref{fig: 2D obs feature space distns}). The very significant overlap in parameter space among the three structural classes, with group galaxies in particular strongly overlapping with main cluster galaxies, suggests that unsupervised clustering analysis in this feature space would not effectively separate the populations relevant to our study. For this reason, we opted for a supervised Random Forest approach, which leverages feature-by-feature distinctions to achieve better classification performance.

\section{ROC and PR curves}\label{sec: appendix B}

\begin{figure*}
    \centering
    \includegraphics[width=0.9\linewidth]{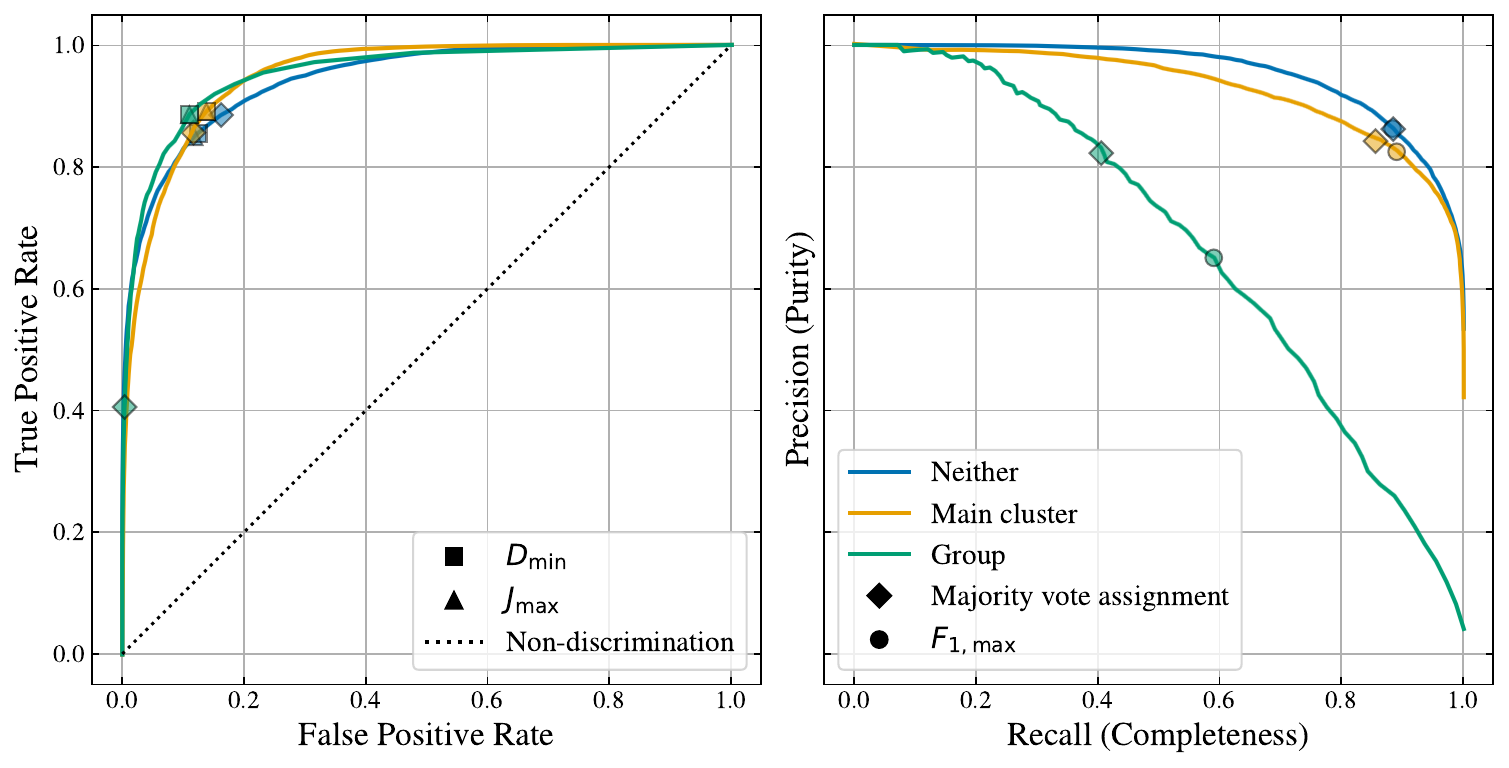}
    \caption{ROC (\textit{left}) and PR (\textit{right}) curves for each class (indicated by colour) using the RF model trained on 3D simulation data and predicted on corresponding validation clusters. The black dotted line in the ROC plot denotes the performance of a classifier no better than random guessing. The diamond markers indicate the point corresponding to the highest-probability (majority vote) prediction scheme. Additional markers show the locations of the optimal scores for the associated performance metrics, serving as reference points to assess how the unoptimised model compares.}
    \label{fig: 3D sim ROC and PR plots}
\end{figure*}

In Figure~\ref{fig: 3D sim ROC and PR plots}, we present Receiver Operating Characteristic (ROC) and Precision-Recall (PR) curves for the RF model trained on 3D-derived feature-space and predicted on the 3D validation clusters

\begin{table*}
    \centering
    \caption{Performance metrics for each class based on ROC and PR curves for 3D validation clusters. When two values are listed per metric, ''Default" values are based on the highest class probability assignment; otherwise, values are computed at the threshold that optimises the corresponding metric.}
    \label{tab:performance-metrics-3D}
    \begin{tabular}{lccccccccccc}
        \toprule
        \textbf{Class} & AUC$_\mathrm{ROC}$ &
        \multicolumn{2}{c}{$J$} &
        \multicolumn{2}{c}{$D$} &
        Pur. & Comp. &
        AUC$_\mathrm{PR}$ & AP &
        \multicolumn{2}{c}{F$_{1}$} \\
        \cmidrule(lr){3-4}
        \cmidrule(lr){5-6}
        \cmidrule(lr){11-12}
         & & Default & Max & Default & Min & & & & & Default & Max \\
        \midrule
        Neither & 0.945 & 0.723 & 0.732 & 0.199 & 0.191 & 0.863 & 0.886 & 0.955 & 0.953  & 0.874 & 0.874 \\
        \midrule
        Main Cluster & 0.950 & 0.740 & 0.753 & 0.185 & 0.176 & 0.843 & 0.857 & 0.929 & 0.928 & 0.850 & 0.857 \\
        \midrule
        Group  & 0.953 & 0.402 & 0.776 & 0.594 & 0.158 & 0.823 & 0.406 & 0.675 & 0.669 & 0.544 & 0.619 \\
        \bottomrule
    \end{tabular}
\end{table*}

The ROC curve depicts the trade-off between a model’s True Positive Rate (TPR; known as sensitivity) and False Positive Rate (FPR) across classification probability thresholds ranging from 0 to 1. At each threshold, a galaxy is assigned to the positive class if its predicted probability exceeds the threshold, even if it has a higher likelihood for another class. Essentially, the ROC curve reflects how well the model distinguishes positives from negatives by maximising true positives while minimising false positives. These are inherently designed for binary classification but are extended to multi-class data by treating each class in a one-vs-rest fashion. For example, when evaluating performance for the group class, we consider predictions of ''group" as the positive class and all other predictions (''main cluster" and ''neither") as negative.

We compute three diagnostic metrics to evaluate model performance: the area under the ROC curve (AUC), Youden’s $J$ index, and the distance to the point (0, 1) on the ROC curve, $D$. The AUC quantifies overall classification performance, representing the probability that a randomly chosen positive instance is ranked higher than a negative one; a value of 0.5 corresponds to random guessing, while 1.0 indicates perfect classification. Youden’s $J$ index is defined as $\mathrm{TPR}-\mathrm{FPR}$ and identifies the threshold that optimally balances sensitivity and specificity. In contrast, $D$ measures the Euclidean distance from each point on the ROC curve to the ideal point (0,1), with the minimum value corresponding to the threshold closest to perfect sensitivity and specificity.

The supplementary PR curve illustrates the relationship between purity (precision; $y$-axis) and completeness (recall; $x$-axis) as a function of probability threshold. Like the ROC curve, model performance can be summarised by the AUC. Another measure is the average purity (AP), which represents the sum of purity values, $P_n$, at each threshold $n$, weighted by the change in completeness between successive thresholds. Higher AP values indicate better performance. An ideal classifier experiences no loss in either purity or completeness, with its PR curve passing through the point (1, 1), achieving an AUC and AP of 1.

Performance metrics for each class, derived from the ROC and PR curves, are summarised in Table~\ref{tab:performance-metrics-3D}. ROC AUC scores are uniformly high across classes (0.94–0.95), indicating strong overall performance. The group class exhibits the highest possible Youden index ($J_{\mathrm{max}}=0.78$), reflecting the greatest separation from the random chance line, and the lowest possible distance to the ideal point on the ROC curve ($D_{\mathrm{min}}=0.16$), signifying high sensitivity. However, group predictions based solely on majority voting remain far from these ideal thresholds. In contrast, the neither and main cluster predictions lie closer to optimal performance, a trend echoed in the PR curves.

The PR AUC and AP for the group class are notably lower (around 0.67) compared to 0.95 and 0.93 for neither and main cluster, respectively. This decline reflects the group class’s minority status. The F$_1$-score for the group class is 0.54 with a ceiling of 0.61. The near-linear group PR curve indicates that threshold adjustment yields limited gains. While optimising for purity enhances the reliability of identified groups, it inevitably reduces recall. As shown in Figure~\ref{fig: cluster 155 yz plane 3d}, many of the retained predictions correspond to group centrals, demonstrating the model’s capacity to isolate high-confidence candidates.

\section{GMM comparison in further detail}\label{sec: appendix C}

\begin{figure}
    \centering
    \includegraphics[width=0.95\linewidth]{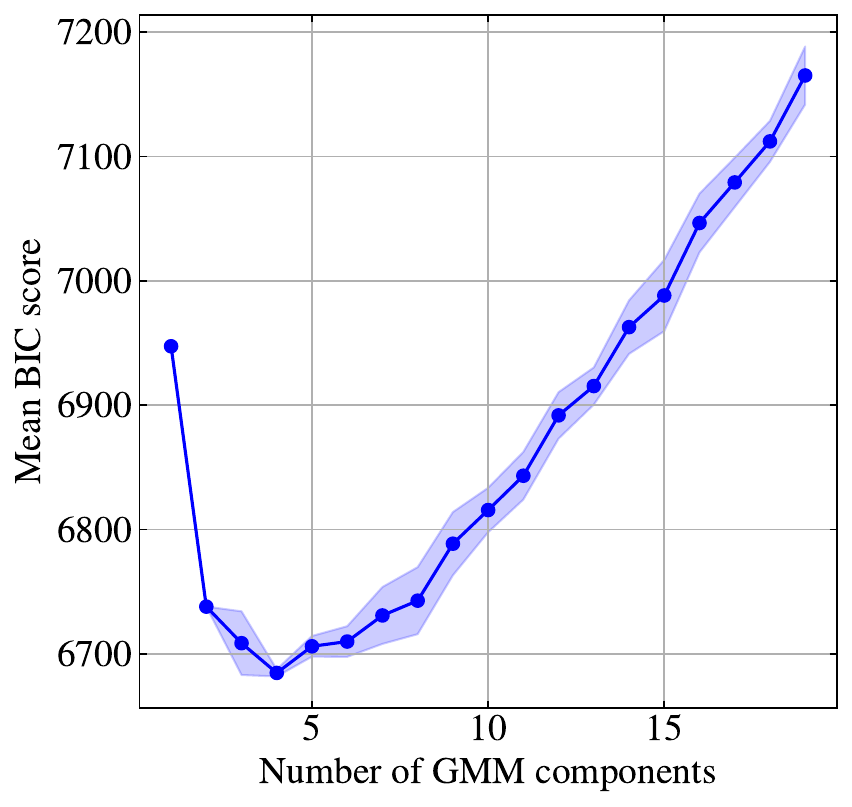}
    \caption{Mean BIC scores for models with varying numbers of Gaussian components fitted to cluster 25 in the $x$–$y$ plane, including velocity data. The solid line indicates the mean value over 10 runs, and the shaded region shows the 1$\sigma$ scatter. The optimal model is with 4 components as this lies at the minimum of the curve.}
    \label{fig:mean_BIC_scores}
\end{figure}

The GMM algorithm is inherently stochastic, so each fitting routine was repeated 10 times for every candidate number of components. The optimal model was identified as the one with the lowest mean BIC score. An example curve for a representative cluster from the test set is shown in Figure~\ref{fig:mean_BIC_scores}.

\begin{figure*}
    \centering
    \includegraphics[width=1\linewidth]{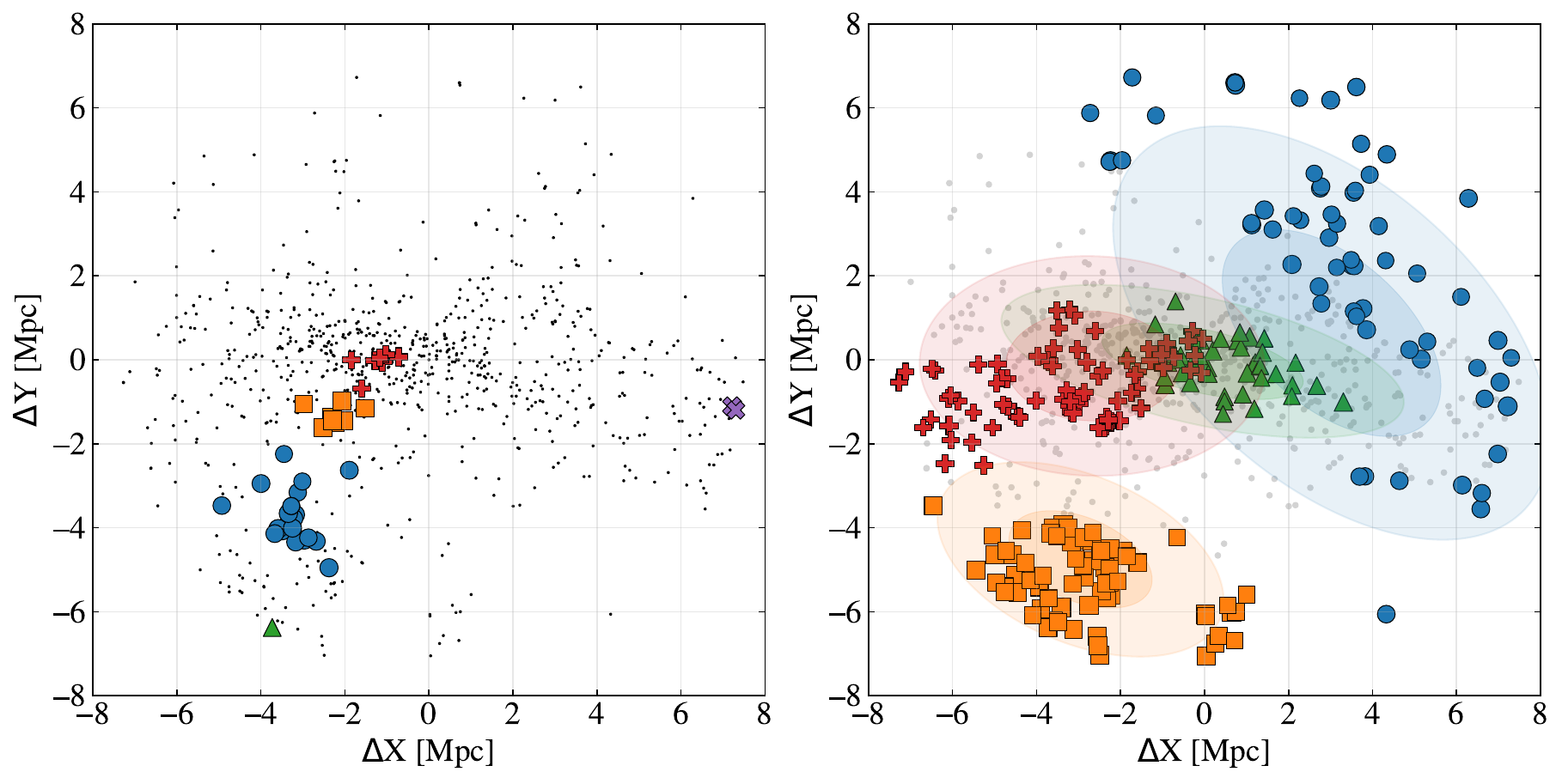}
    \caption{Gaussian Mixture Modelling results for cluster 25 in the $x$–$y$ sky-plane projection. Left: true galaxy groups, distinguished by colour and marker. Right: the optimal GMM solution with four components, where galaxies are assigned to coloured components (colours are not related to the true groups). Galaxies with membership probability below 97\% are shown as grey background points. Ellipses mark the fitted Gaussian contours at 1$\sigma$ (darker) and 2$\sigma$ (lighter), indicating the extent of the components.}
    \label{fig:cluster_25_xy_GMM_results}
\end{figure*}

A clear minimum occurs at four components, even accounting for scatter, and this result is adopted as the optimal model. The BIC statistic balances model likelihood against complexity (i.e. the number of parameters). As expected, BIC increases steadily with larger numbers of components: additional Gaussians can always provide a tighter fit to the data, but at the cost of overfitting. This trade-off is an inherent limitation of the method. An example of the resulting clustering for the optimal model is shown in Figure~\ref{fig:cluster_25_xy_GMM_results}, where galaxy membership has been refined using a 97\% probability threshold.

\begin{figure}
    \centering
    \includegraphics[width=\linewidth]{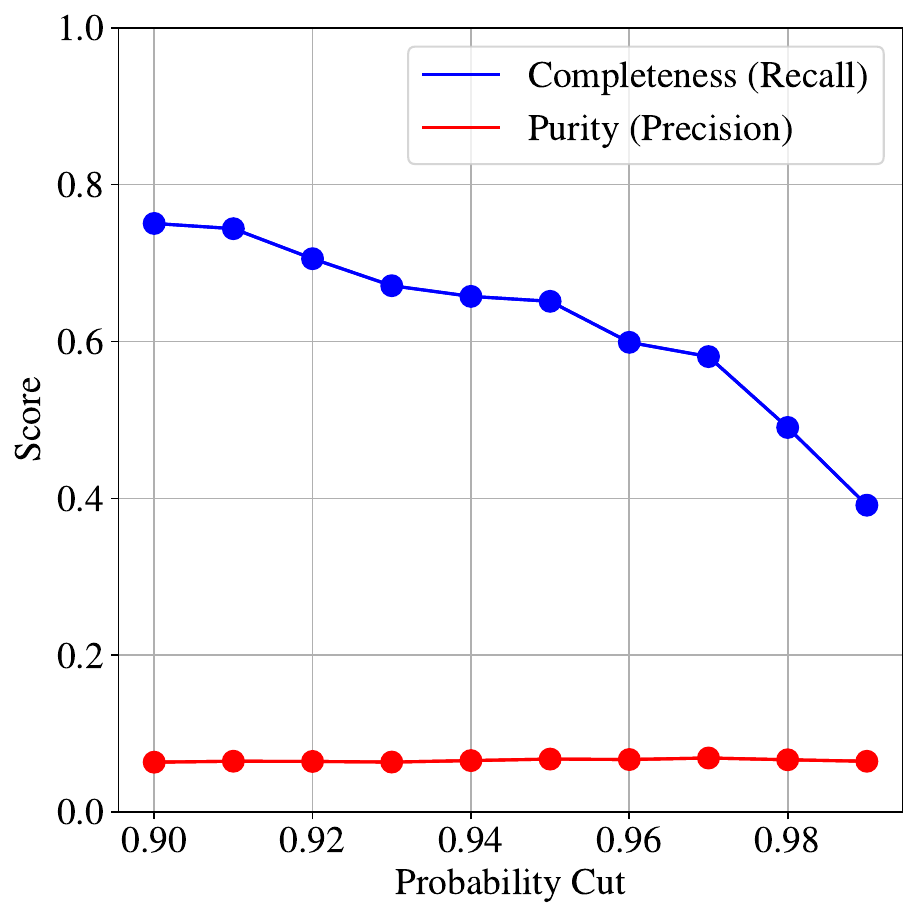}
    \caption{Variation of completeness and purity with probability threshold for GMM applied across all test cluster projections.}
    \label{fig:c_and_p_vs_prob_cut}
\end{figure}

The GMM approach consistently recovers dense structures in cluster cores, although the identified components do not necessarily correspond to the true groups. Even at a stringent probability cut, components remain rich in membership: for this projection, completeness is boosted to 67\% but purity remains low at 10\%. Similar behaviour is observed across all cluster projections. To quantify this trend, we applied GMM to all test cluster projections and measured completeness and purity as a function of probability threshold. The resulting curves are shown in Figure~\ref{fig:c_and_p_vs_prob_cut}.

These results demonstrate that tightening the membership criterion will marginally reduce completeness but does not enhance purity, which remains fundamentally low.


\bsp
\label{lastpage}
\end{document}